\documentclass[%
 aip,
 amsmath,amssymb,
 reprint,%
]{revtex4-1}

\usepackage{graphicx}
\usepackage{dcolumn}
\usepackage{bm}

\usepackage[utf8]{inputenc}
\usepackage[T1]{fontenc}
\usepackage{mathptmx}
\usepackage{etoolbox}
\usepackage{xcolor}

\usepackage{array}
\newcommand{\PreserveBackslash}[1]{\let\temp=\\#1\let\\=\temp}
\newcolumntype{C}[1]{>{\PreserveBackslash\centering}p{#1}}
\newcolumntype{R}[1]{>{\PreserveBackslash\raggedleft}p{#1}}
\newcolumntype{L}[1]{>{\PreserveBackslash\raggedright}p{#1}}

\makeatletter
\def\@email#1#2{%
 \endgroup
 \patchcmd{\titleblock@produce}
  {\frontmatter@RRAPformat}
  {\frontmatter@RRAPformat{\produce@RRAP{*#1\href{mailto:#2}{#2}}}\frontmatter@RRAPformat}
  {}{}
}%
\makeatother
\begin{document}

\preprint{AIP/123-QED}

\title{Triple electron-electron-proton excitations and second-order approximations in nuclear-electronic orbital coupled cluster methods} 

\author{Fabijan Pavo\v{s}evi\'{c}}
\email{fpavosevic@gmail.com}
\affiliation{Center for Computational Quantum Physics, Flatiron Institute, 162 5th Ave., New York, 10010  NY,  USA}

\author{Sharon Hammes-Schiffer}
\affiliation{Department of Chemistry, Yale University, 225 Prospect Street, New Haven, Connecticut, 06520, USA}


\begin{abstract}
The accurate description of nuclear quantum effects, such as zero-point energy, is important for modeling a wide range of chemical and biological processes. Within the nuclear-electronic orbital (NEO) approach, such effects are incorporated in a computationally efficient way by treating electrons and select nuclei, typically protons, quantum mechanically with molecular orbital techniques. Herein, we implement and test a NEO coupled cluster method that explicitly includes the triple electron-proton excitations, where two electrons and one proton are excited simultaneously. Our calculations show that this NEO-CCSD(eep) method provides highly accurate proton densities and proton affinities, outperforming any previously studied NEO method. These examples highlight the importance of the triple electron-electron-proton excitations for an accurate description of nuclear quantum effects. Additionally, we also implement and test the second-order approximate coupled cluster with singles and doubles (NEO-CC2) method, as well as its scaled-opposite-spin (SOS) versions. 
The NEO-SOS$'$-CC2 method, which scales the electron-proton correlation energy as well as the opposite-spin and same-spin components of the electron-electron correlation energy, achieves nearly the same accuracy as the NEO-CCSD(eep) method for the properties studied. Because of its low computational cost, this method will enable a wide range of chemical and photochemical applications for large molecular systems. This work sets the stage for a wide range of developments and applications within the NEO framework.

\end{abstract}

\maketitle

\section{Introduction}
Multicomponent quantum chemistry methods, in which more than one type of particle (e.g., electrons, positrons, nuclei, or photons) is treated quantum mechanically, are promising theoretical tools for describing various types of interesting chemical phenomena.~\cite{pavosevic2020chemrev,ruggenthaler2018quantum,pavosevic2021cavity} Among different multicomponent approaches,\cite{ishimoto2009review,pavosevic2020chemrev,reyes2019any} the nuclear-electronic orbital (NEO) method~\cite{webb2002multiconfigurational,pavosevic2020chemrev} treats all electrons and specified nuclei, typically protons, quantum mechanically on the same footing with molecular orbital techniques. In this way, many important nuclear quantum effects, such as zero-point energy, proton delocalization, and hydrogen tunnelling, as well as non-Born-Oppenheimer effects, are included during energy and reaction path calculations in a computationally efficient way.  

The simplest method that can be formulated within the NEO framework is NEO-Hartree-Fock (NEO-HF),~\cite{webb2002multiconfigurational} in which the wave function is represented as a direct product of single electronic and single protonic Slater determinants. Because the NEO-HF method treats electrons and protons as uncorrelated particles, the predictions obtained from this method, such as proton densities and resulting properties, are highly inaccurate and unreliable.~\cite{pavosevic2018ccsd,pavovsevic2019multicomponent,pavosevic2020chemrev} 
Analogous to conventional electronic structure methods, there are two main strategies to incorporate the missing correlation effects between quantum particles: density functional theory (DFT)~\cite{pak2007density,yang2017development,brorsen2017multicomponent} and wave function theory.~\cite{webb2002multiconfigurational,nakai2003many,ishimoto2009review,pavosevic2018ccsd,pavovsevic2019multicomponent,pavosevic2020multicomponent,pavosevic2021dfccsd,fajen2020separation,fajen2021multicomponent} In the NEO-DFT method, both electron-electron and electron-proton correlation effects are included via correlation functionals in a computationally practical manner.\cite{yang2017development,brorsen2017multicomponent} Because this method balances accuracy and computational cost, it is suitable for treatment of large molecular systems. However, a disadvantage of the NEO-DFT method is that it is not systematically improvable, and it suffers from the same problems that are inherent to conventional DFT methods,\cite{hermann2017} such as self-interaction error.~\cite{cohen2008insights}

As an alternative to NEO-DFT, the wave function based methods, such as the NEO coupled cluster (NEO-CC) methods,\cite{nakai2003many,ellis2016development,pavosevic2018ccsd,pavovsevic2019multicomponent,pavosevic2020multicomponent,pavosevic2021dfccsd,pavosevic2021multicomponent} are systematically improvable and parameter free. The NEO-CC methods use the exponentiated cluster operator to incorporate the correlation effects between quantum particles (i.e., electrons and protons) via single, double, and higher excitation ranks.\cite{crawford2000introduction,bartlett2007coupled,shavitt2009many} The truncation of the cluster operator up to a certain excitation rank establishes the NEO-CC hierarchy. For example, truncation of the cluster operator to include up to single and double electronic and protonic excitations, as well as double electron-proton excitations, defines the NEO coupled cluster with singles and doubles (NEO-CCSD) method.\cite{pavosevic2018ccsd} Previously we showed that the NEO-CCSD method accurately predicts proton densities, energies, and vibrationally averaged geometries.\cite{pavosevic2018ccsd,pavovsevic2019multicomponent} More recently, the computational efficiency of the NEO-CCSD method was enhanced by the density fitting (DF) scheme,\cite{pavosevic2021dfccsd} which significantly reduces the memory requirements. This strategy enabled calculations of proton affinities of much larger molecules than previously possible, as well as the study of relative stabilities of protonated water tetramers with all nine protons treated quantum mechanically.\cite{pavosevic2021dfccsd} The reliability and robustness of the NEO-CCSD method has sparked an interest in development of other NEO wave function based methods, most notably the computationally attractive scaled-opposite-spin orbital optimized second-order M{\o}ller-Plesset perturbation theory (NEO-SOS$'$-OOMP2) method,\cite{pavosevic2020multicomponent,fetherolf_oomp2} which scales the electron-proton correlation energy as well as the opposite-spin components of the electron-electron correlation energy.  

In the NEO-CCSD method, truncation of the cluster operator to include up to double electron-proton excitations represented a compromise between accuracy and computational efficiency, as well as simplicity of implementation.\cite{pavosevic2018ccsd} In order to account for some of the missing electron-proton correlation, in our previous work we used a larger electronic basis set for the quantum proton(s) than for the other nuclei.\cite{pavosevic2018ccsd} Although this strategy works well by providing accurate predictions of different properties for the studied systems,\cite{pavosevic2018ccsd,pavovsevic2019multicomponent,pavosevic2021dfccsd} it might not be general for all systems. In this work, we move beyond the NEO-CCSD method by implementing the NEO-CCSD(eep) method, which also includes electron-electron-proton triple excitations. The importance of such triple excitations was observed recently in the context of perturbation theory.\cite{fajen2021mp4} Additionally, we implement and investigate a novel and computationally efficient second-order approximate coupled cluster with singles and doubles (NEO-CC2) method and its scaled-opposite-spin version (NEO-SOS$'$-CC2). Analogous to its electronic counterpart,\cite{christiansen1995second,hellweg2008benchmarking,tajti2019accuracy} the NEO-CC2 method can be used as a computationally efficient alternative to the equation-of-motion coupled cluster methods for excited states.\cite{pavovsevic2019eomccsd,pavosevic2020frequency} Moreover, in order to calculate protonic densities with these methods, we also implement
the $\Lambda$-equations using automatic differentiation.\cite{pavovsevic2020automatic} The developments and tests performed within this work highlight the robustness and reliability of the NEO-CC methods. 

\section{Theory}
In this section we describe the multicomponent wave function approaches in which electrons and protons are treated quantum mechanically. We note that the extension to other multicomponent fermionic systems, such as where positrons instead of protons are treated quantum mechanically, is straightforward.\cite{pavosevic2020chemrev}

The NEO coupled cluster correlation energy is calculated from the energy Lagrangian as\cite{pavovsevic2019multicomponent}
\begin{equation}
    \label{eqn:NEO-CC-Lagrangian}
    E_{\text{NEO-CC}}=\langle0^{\text{e}}0^{\text{p}}|(1+\hat{\Lambda})e^{-\hat{T}}\hat{H}_{\text{NEO}}e^{\hat{T}}|0^{\text{e}}0^{\text{p}}\rangle.
\end{equation}
In this equation, $\hat{H}_\text{NEO}$ is the second-quantized normal-ordered (with respect to the NEO-HF reference state, $|0^{\text{e}}0^{\text{p}}\rangle$) NEO Hamiltonian that is expressed as
\begin{equation}
    \label{eqn:NEO-Hamiltonian}
    \hat{H}_\text{NEO} = F^p_q a^q_p + \frac{1}{4}\overline{g}^{pq}_{rs}a^{rs}_{pq}+F^P_Q a^Q_P + \frac{1}{4}\overline{g}^{PQ}_{RS}a^{RS}_{PQ} - g^{pP}_{qQ} a^{qQ}_{pP},
\end{equation}
where $a_{p_1p_2...p_n}^{q_1q_2...q_n}=a_{q_1}^{\dagger}a_{q_2}^{\dagger}...a_{q_n}^{\dagger}a_{p_n}...a_{p_2}a_{p_1}$ are normal-ordered second-quantized excitation operators written in terms of fermionic creation/annihilation ($a^{\dagger}/a$) operators. The lowercase indices $i,j,k,l,...$, $a,b,c,d,...$, and $p,q,r,s,...$ denote occupied, unoccupied, and general electronic spin orbitals, whereas the corresponding uppercase indices denote protonic orbitals. Additionally, $F^p_q=\langle q |\hat{F}^\text{e}|p \rangle$ is a matrix element of
the electronic Fock operator and $\overline{g}^{pq}_{rs}=g^{pq}_{rs}-g^{qp}_{rs}=\langle rs|pq \rangle - \langle rs|qp \rangle$ is an antisymmetrized two-electron repulsion tensor element. Their protonic counterparts $F^P_Q$ and $\overline{g}^{PQ}_{RS}$ are defined analogously, and
$g^{pP}_{qQ}=\langle qQ|pP \rangle$ is the electron-proton attraction tensor element. The Einstein summation convention over repeated indices is utilized throughout this manuscript. 

In Eq. \eqref{eqn:NEO-CC-Lagrangian}, $\hat{T}=t_{\mu}a^{\mu}$ and $\hat{\Lambda}=\lambda^{\mu}a_{\mu}$ are excitation and de-excitation cluster operators, respectively, where $a^{\mu}=a_{\mu}^{\dagger}=\{a_{i}^{a},a_{I}^{A},a_{iI}^{aA},a_{ij}^{ab},a_{IJ}^{AB},a_{ijI}^{abA},a_{iIJ}^{aAB},...\}$ is a set of single, double, and higher excitation operators, and $\mu$ is an excitation rank. Moreover, $t_{\mu}$ and $\lambda^{\mu}$ are unknown wave function parameters (amplitudes) that are determined by minimizing Eq. \eqref{eqn:NEO-CC-Lagrangian} with respect to $\lambda^{\mu}$ and $t_{\mu}$, respectively:
\begin{equation}
    \label{eqn:t-equations}
    \frac{\partial E_{\text{NEO-CC}}}{\partial \lambda^{\mu}}=\langle0^{\text{e}}0^{\text{p}}|a_{\mu}e^{-\hat{T}}\hat{H}_{\text{NEO}}e^{\hat{T}}|0^{\text{e}}0^{\text{p}}\rangle=0,
\end{equation}
\begin{equation}
    \label{eqn:lambda-equations}
    \frac{\partial E_{\text{NEO-CC}}}{\partial t_{\mu}}=\langle0^{\text{e}}0^{\text{p}}|(1+\hat{\Lambda})[e^{-\hat{T}}\hat{H}_{\text{NEO}}e^{\hat{T}},a^{\mu}]|0^{\text{e}}0^{\text{p}}\rangle=0.
\end{equation}
The last two equations are known as the $t$-amplitude equations and the $\Lambda$-equations, respectively. The truncation of the cluster operator $\hat{T}$ up to a certain excitation rank $\mu$ establishes the NEO-CC hierarchy. 

In our previous work,\cite{pavosevic2018ccsd,pavovsevic2019multicomponent,pavosevic2021dfccsd} the cluster operator $\hat{T}$ was defined as
\begin{equation}
    \label{eqn:tep}
    \hat{T}^{\text{(ep)}}=t^{i}_{a}a_{i}^{a}+t^{I}_{A}a_{I}^{A}+t^{iI}_{aA}a_{iI}^{aA}+\frac{1}{4}t^{ij}_{ab}a_{ij}^{ab}+\frac{1}{4}t^{IJ}_{AB}a_{IJ}^{AB}.
\end{equation}
Because the highest level of electron-proton excitation is the simultaneous single electronic and single protonic excitations due to $a_{iI}^{aA}$, we will refer to this method as NEO-CCSD(ep) throughout this manuscript. In the present work, we implement and explore the NEO-CC method with the cluster operator defined as
\begin{equation}
    \label{eqn:teep}
    \hat{T}^{\text{(eep)}}=t^{i}_{a}a_{i}^{a}+t^{I}_{A}a_{I}^{A}+t^{iI}_{aA}a_{iI}^{aA}+\frac{1}{4}t^{ij}_{ab}a_{ij}^{ab}+\frac{1}{4}t^{IJ}_{AB}a_{IJ}^{AB}+\frac{1}{4}t^{ijI}_{abA}a_{ijI}^{abA}.
\end{equation}
This cluster operator explicitly includes simultaneous double electronic and single protonic excitations. Although the total excitation rank of the operator $a_{ijI}^{abA}$ is triple, the highest excitation rank of a single  particle (in this case electrons) is double, and therefore this method is denoted NEO-CCSD(eep). Although the addition of one extra term into the cluster operator may seem to be a trivial extension, the $t$-amplitude equations of the new NEO-CCSD(eep) method have roughly four times more terms than the NEO-CCSD(ep) method. Therefore, the derivation and implementation of the working equations for the NEO-CCSD(eep) method require a significant amount of effort. The computational cost of the NEO-CCSD(ep) and NEO-CCSD(eep) methods scales as $\mathcal{O}(N^6)$, where $N$ is a measure of the system size, although the NEO-CCSD(eep) method has a greater prefactor than the NEO-CCSD(ep) method. For electron-dominated systems with one quantum-proton (as considered in the present study), the majority of the computation time for the NEO-CCSD(ep) method is spent in determining the $t^{ij}_{ab}$ amplitudes. On the other hand, the NEO-CCSD(eep) method has an additional set of $t$-amplitude equations for determining the $t^{ijI}_{abA}$ amplitudes. The total cost for the NEO-CCSD(eep) method is expressed roughly as the cost of determining the $t^{ij}_{ab}$ amplitudes multiplied by the number of protonic basis functions. 

The programmable expressions for the $t$-amplitude equations of the NEO-CCSD(eep) method are obtained by utilizing the generalized Wick’s theorem.\cite{kutzelnigg1997normal,shavitt2009many,bartlett2007coupled} The $\Lambda$-equations can in principle be derived in the same way,\cite{pavovsevic2019multicomponent} but because they have 50\% more terms than the $t$-amplitude equations,\cite{pavovsevic2019multicomponent,pavovsevic2020automatic} their derivation and implementation is a daunting task. Alternatively, the unknown $\lambda^{\mu}$ amplitudes can be calculated with the aid of automatic differentiation, as illustrated in our previous work.\cite{pavovsevic2020automatic} Within this procedure, the Lagrangian given in Eq. \eqref{eqn:NEO-CC-Lagrangian} is constructed by augmenting the NEO-CC energy $\big(\langle0^{\text{e}}0^{\text{p}}|e^{-\hat{T}}\hat{H}_{\text{NEO}}e^{\hat{T}}|0^{\text{e}}0^{\text{p}}\rangle\big)$ with the $t$-amplitude equations from Eq. \eqref{eqn:t-equations} weighted by the Lagrange multipliers $\lambda^{\mu}$. Therefore, if the $t$-amplitude equations are available, construction of the Lagrangian is straightforward. Once the Lagrangian is available, both $t_{\mu}$ and $\lambda^{\mu}$ are calculated with automatic differentiation. A significant advantage of this procedure is that it does not require derivation and implementation of the $\Lambda$-equations, thereby immensely reducing the coding effort. 

The calculated wave function parameters $t_{\mu}$ and $\lambda^{\mu}$ allow calculation of various important molecular properties, one of which is the proton density that is used to validate the accuracy of NEO methods.\cite{pavovsevic2019multicomponent} Accurate proton densities are crucial for calculation of molecular properties, such as vibrationally averaged geometries and zero-point energies.\cite{brorsen2017multicomponent,pavosevic2020chemrev} The proton density is calculated from
\begin{equation}
    \label{eqn:density}
    \rho_{\text{p}}(\textbf{r}_{\text{p}})=\sum_{PQ}\gamma_{P}^{Q}\phi_P(\textbf{r}_{\text{p}})\phi_Q(\textbf{r}_{\text{p}}),
\end{equation}
where $\gamma_{P}^{Q}$ is the total one-particle reduced density matrix that is defined as $\gamma_{P}^{Q}=\gamma_{\text{NEO-HF}}+\tilde{\gamma}_{P}^{Q}$. Here, $\gamma_{\text{NEO-HF}}$ is the NEO-HF one-particle reduced density matrix, and $\tilde{\gamma}_{P}^{Q}$ is the NEO-CC one-particle reduced density matrix defined by
\begin{equation}
    \label{eqn:1rdm}
    \tilde{\gamma}_{P}^{Q}=\langle0^{\text{e}}0^{\text{p}}|(1+\Lambda)e^{-\hat{T}}a_P^Qe^{\hat{T}}|0^{\text{e}}0^{\text{p}}\rangle.
\end{equation}
In Eq. \eqref{eqn:density}, $\phi_{P}$ is a protonic orbital and $\textbf{r}_{\text{p}}$ is the proton coordinate.

In this work, we also explore the second-order coupled cluster (NEO-CC2) method within the NEO framework, which can be regarded as an approximate NEO-CCSD(ep) method. In the NEO-CC2 method, the singles $t$-amplitude equations remain the same and are equivalent to those of the NEO-CCSD(ep) method, whereas the doubles $t$-amplitude equations are approximated as
\begin{equation}
    \label{eqn:t-equations-cc2}
    \langle0^{\text{e}}0^{\text{p}}|a_{\mu_2}\big(\bar{H}+[\hat{F},\hat{T}_2]\big)|0^{\text{e}}0^{\text{p}}\rangle=0.
\end{equation}
Here, $a_{\mu_2}=\{a^{iI}_{aA},a^{ij}_{ab},a^{IJ}_{AB}\}$ is the double de-excitation operator, $\bar{H}=e^{-\hat{T}_1}\hat{H}_{\text{NEO}}e^{\hat{T}_1}$ is the normal-ordered $\hat{T}_1$-similarity transformed NEO Hamiltonian, and $\hat{F}=F^p_q a^q_p + F^P_Q a^Q_P$ is the normal-ordered second-quantized Fock operator. The cluster operators used in this expression are defined as
\begin{equation}
    \label{eqn:t1-cc2}
    \hat{T}_1=t^{i}_{a}a_{i}^{a}+t^{I}_{A}a_{I}^{A}
\end{equation}
and
\begin{equation}
    \label{eqn:t2-cc2}
    \hat{T}_2=t^{iI}_{aA}a_{iI}^{aA}+\frac{1}{4}t^{ij}_{ab}a_{ij}^{ab}+\frac{1}{4}t^{IJ}_{AB}a_{IJ}^{AB}.
\end{equation}
Due to the approximations introduced in the NEO-CC2 method, the computational cost scales as $\mathcal{O}(N^5)$.  

The NEO-CC2 method is closely related to the NEO-MP2 and NEO-OOMP2 methods. The NEO-MP2 method is obtained by setting the singles amplitudes in the NEO-CC2 method to zero, whereas the NEO-OOMP2 method is obtained by using the unitary rotations of orbitals instead of the exponentiated singles operator. The working equations of the NEO-CC2 and NEO-OOMP2 methods are very similar, as discussed in the context of their purely electronic counterparts in Ref.~\citenum{neese2009assessment}. Calculations of the $\Lambda$-equations and protonic density are performed analogously with the NEO-CC2 method as with the NEO-CCSD(ep) and NEO-CCSD(eep) methods. 

The computational efficiency and accuracy of the NEO-CC2 method can be enhanced with the SOS approach,~\cite{jung2004scaled,distasio2007optimized,hellweg2008benchmarking,tajti2019accuracy} in which the opposite-spin and same-spin components of the electron-electron correlation energy are scaled differently. In the context of the NEO method, the accuracy can be further enhanced by scaling the electron-proton contribution of the correlation energy,\cite{pavosevic2020multicomponent} leading to the NEO-SOS$'$-CC2 method. Within this approach, the working singles and doubles amplitude equations are modified as follows:
\begin{equation}
    \label{eqn:t1-equations-sos-cc2}
    \langle0^{\text{e}}0^{\text{p}}|a_{\mu_1}\big(\bar{H}+\sum_{\sigma,\sigma'}c_{\sigma,\sigma'}[\bar{H},\hat{T}_{2,\text{ee}}^{\sigma,\sigma'}]+c_{\text{ep}}[\bar{H},\hat{T}_{2,\text{ep}}]\big)|0^{\text{e}}0^{\text{p}}\rangle=0,
\end{equation}
\begin{equation}
    \label{eqn:t2-equations-sos-cc2}
    \langle0^{\text{e}}0^{\text{p}}|a_{\mu_2}\big(\bar{H}+\sum_{\sigma,\sigma'}c_{\sigma,\sigma'}[\hat{F},\hat{T}_{2,\text{ee}}^{\sigma,\sigma'}]+c_{\text{ep}}[\hat{F},\hat{T}_{2,\text{ep}}]\big)|0^{\text{e}}0^{\text{p}}\rangle=0,
\end{equation}
respectively. The NEO-SOS$'$-CC2 energy is calculated from
\begin{equation}
    \label{eqn:sos-cc2-energy}
    \begin{split}
    E_{\text{NEO-SOS$'$-CC2}}=&\langle0^{\text{e}}0^{\text{p}}|\big(\bar{H}+\sum_{\sigma,\sigma'}c_{\sigma,\sigma'}[\bar{H},\hat{T}_{2,\text{ee}}^{\sigma,\sigma'}]\\&+c_{\text{ep}}[\bar{H},\hat{T}_{2,\text{ep}}]\big)|0^{\text{e}}0^{\text{p}}\rangle.
    \end{split}
\end{equation}
In the last three equations, $\sigma$/$\sigma'$ indicates $\alpha$/$\beta$ electron spin, $c_{\sigma,\sigma'} = \{ c_{\text{os}}=c_{\alpha\beta}=c_{\beta\alpha},c_{\text{ss}}=c_{\alpha\alpha}=c_{\beta\beta}\}$ are electron spin-specific scaling coefficients, $\hat{T}_{2,\text{ee}}^{\sigma,\sigma'}=\frac{1}{2}t^{i_{\sigma}j_{\sigma'}}_{a_{\sigma}b_{\sigma'}}a_{i_{\sigma}j_{\sigma'}}^{a_{\sigma}b_{\sigma'}}$ is the spin-specific purely electronic cluster operator, $\hat{T}_{2,\text{ep}}=t^{iI}_{aA}a_{iI}^{aA}$ is the electron-proton cluster operator, and $c_{\text{ep}}$ is the scaling coefficient for the electron-proton correlation energy contribution. In the conventional electronic structure SOS-CC2 method, the opposite-spin and the same-spin scaling parameters are $c_{\text{os}}=1.3$ and $c_{\text{ss}}=0.0$, respectively.~\cite{jung2004scaled,tajti2019accuracy} Neglecting the same-spin electron-electron correlation allows implementation of the SOS-CC2 and NEO-SOS$'$-CC2 methods with $\mathcal{O}(N^4)$ scaling.~\cite{jung2004scaled,tajti2019accuracy}

Throughout this work, we apply density fitting~\cite{whitten1973coulombic,dunlap1979some}
to approximate the four-center two-particle integrals from Eq. \eqref{eqn:NEO-Hamiltonian} as\cite{mejia2019multicomponent,pavosevic2021dfccsd}
\begin{subequations}
    \begin{align}
    \begin{split}
    g^{pq}_{rs} &= (rp|sq) \\ &\approx \sum_{MN}(rp|M)(M|N)^{-1}(N|sq),
    \end{split}
    \\
    \begin{split}
    g^{PQ}_{RS} &= (RP|SQ) \\ &\approx \sum_{M'N'}(RP|M')(M'|N')^{-1}(N'|SQ),
    \end{split}
    \\
    \begin{split}
    g^{pP}_{qQ} & =(pq|QP) \\ &\approx \sum_{M'N'}(pq|M')(M'|N')^{-1}(N'|QP).
    \end{split}
    \label{eq:densefitmixed}
    \end{align}
\end{subequations}
In these equations, $M$ and $N$ indices denote auxiliary electronic basis functions, and $M'$ and $N'$ indices denote auxiliary protonic basis functions. Within the density fitting approach, the four-center two-particle integrals (here expressed in the chemist notation) are approximated in terms of the three-center and two-center two-particle integrals, thereby significantly reducing the memory requirements.

\section{Results}
The NEO-CCSD(ep), NEO-CCSD(eep), NEO-CC2, and NEO-SOS$'$-CC2 methods were implemented in an in-house version of the Psi4NumPy quantum chemistry software.~\cite{smith2018psi4numpy} All the implemented methods rely on the density fitting scheme for approximating the four-center two-particle integrals. The programmable expressions of the $t$-amplitude equations for the NEO-CCSD(eep) method have been derived with the SeQuant software.~\cite{githubsequant} The $\Lambda$-equations were solved using automatic differentiation with the procedure described elsewhere.~\cite{pavovsevic2020automatic} Automatic differentiation was performed using the TensorFlow v2.1.0 program.~\cite{tensorflow2015} The NEO methods were used to calculate proton densities for the FHF$^-$ and HCN molecules, as well as proton affinities for a set of 12 small molecules.~\cite{pavosevic2018ccsd} All of the calculations were performed at the equilibrium geometries optimized with the conventional electronic CCSD/aug-cc-pVTZ level of theory. In the present study, the calculations employed the aug-cc-pVXZ~\cite{dunning1989gaussian,kendall1992electron} electronic basis set along with its matching  aug-cc-pVXZ-RI~\cite{weigend1998ri,weigend2002efficient} electronic auxiliary basis set, where the basis set cardinal number is X=D,T,Q,5. Moreover, the quantum protons were treated with the PB4-F2 (4s3p2d2f) nuclear basis set~\cite{yu2020development} as well with an even-tempered 8s8p8d8f auxiliary nuclear basis set with exponents ranging from 2$\sqrt{2}$ to 32.~\cite{culpitt2019enhancing} The electronic and nuclear basis sets for the quantum hydrogen were centered at the hydrogen position optimized with the CCSD method.

\begin{figure}[ht]
  \centering
  \includegraphics[width=3.25in]{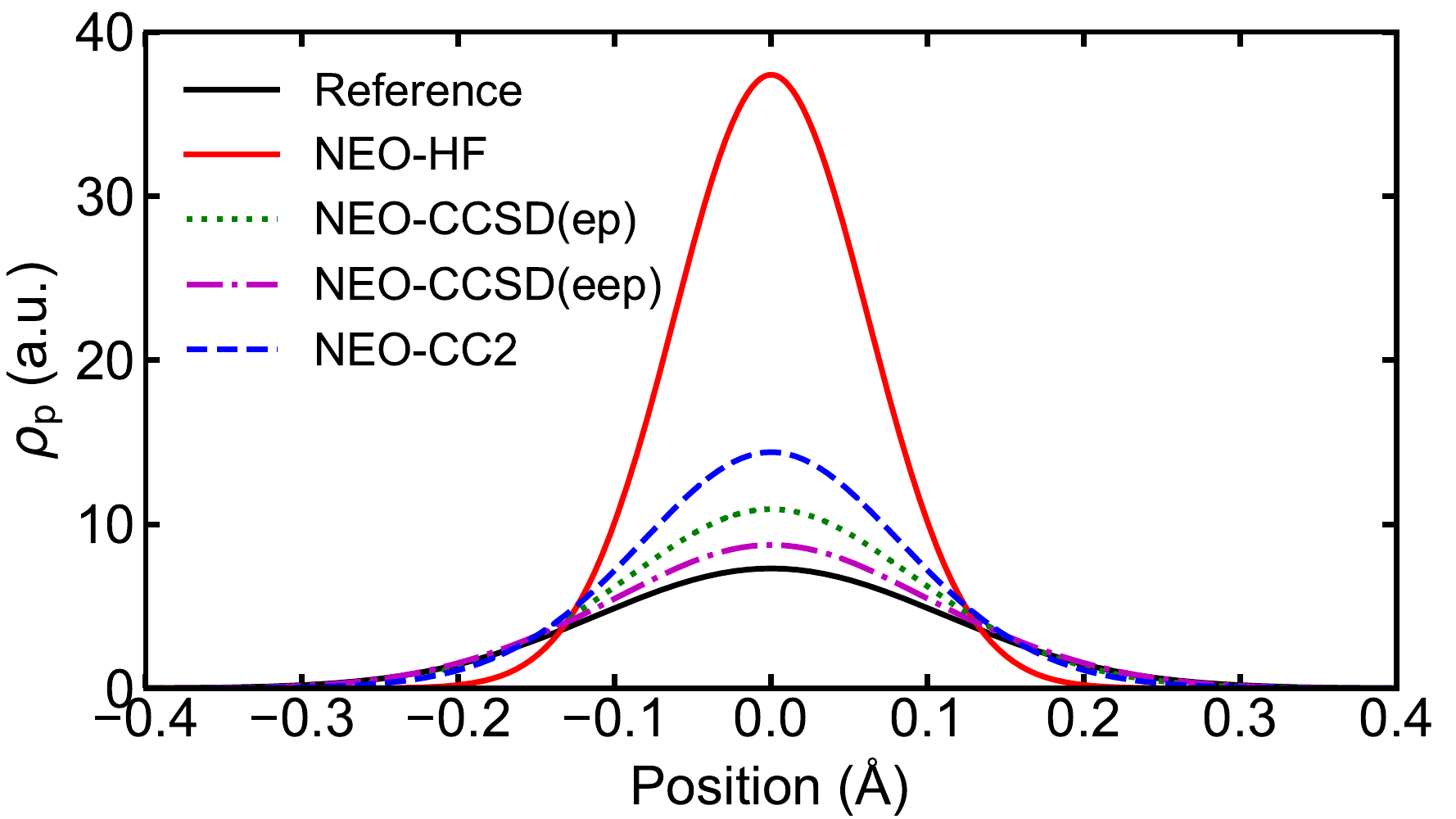}
  \caption{On-axis proton density for the FHF$^-$ molecule calculated with the reference FGH (black solid curve), NEO-HF (red solid curve), NEO-CC2 (blue dashed curve), NEO-CCSD(ep) (green dotted curve), and NEO-CCSD(eep) (mangenta dashed-dotted curve) methods. The NEO calculations employ the aug-cc-pV5Z electronic basis set. The fluorine atoms are positioned at -1.1335~Å and 1.1335~Å, and the proton basis functions are centered at the origin. The on-axis proton density is along the line that connects the two fluorine atoms.}
  \label{fig:fhf_density}
\end{figure}

\begin{figure}[ht]
  \centering
  \includegraphics[width=3.25in]{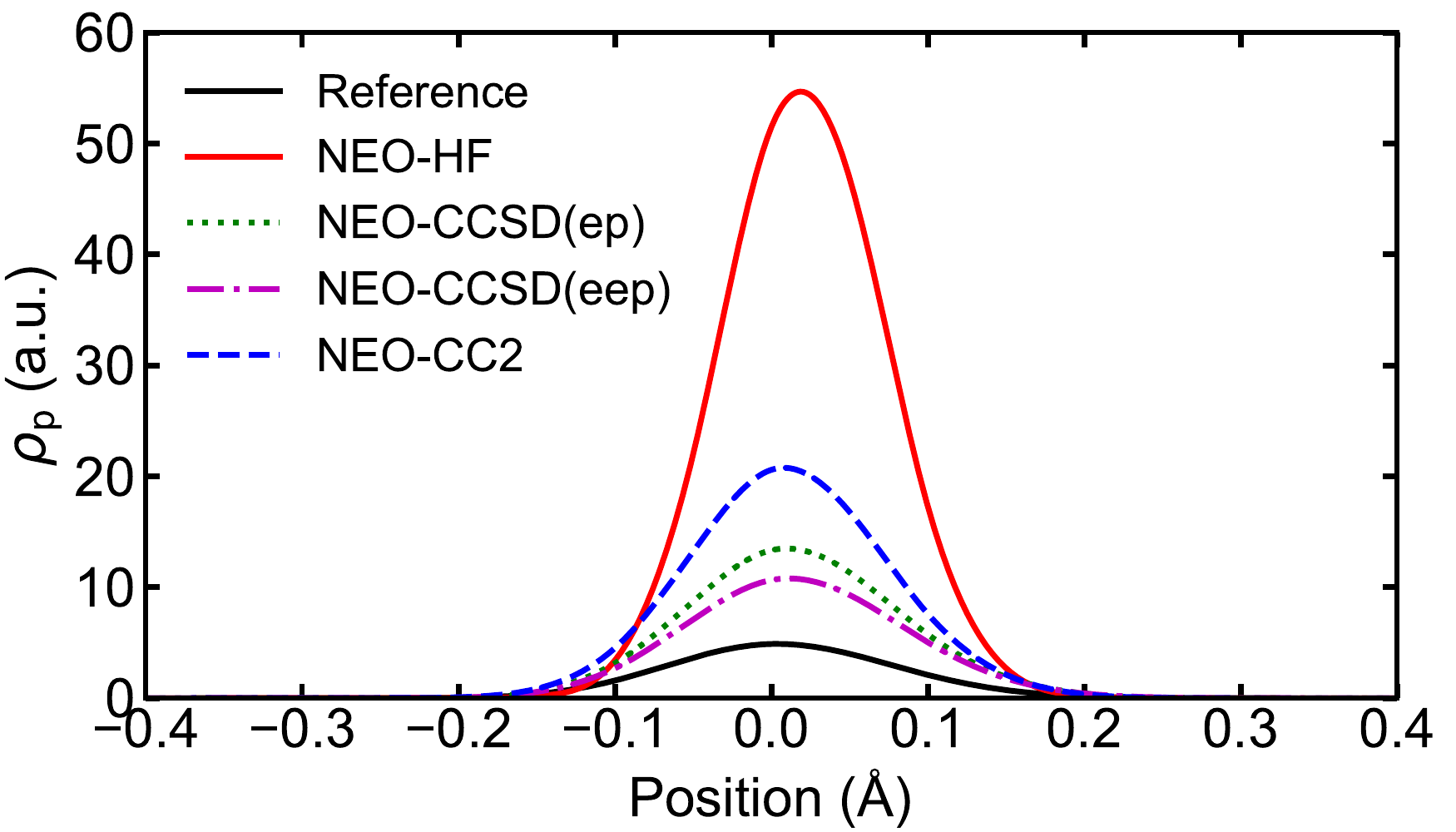}
  \caption{On-axis proton density for the HCN molecule calculated with the reference FGH (black solid curve), NEO-HF (red solid curve), NEO-CC2 (blue dashed curve), NEO-CCSD(ep) (green dotted curve), and NEO-CCSD(eep) (mangenta dashed-dotted curve) methods. The NEO calculations employ the aug-cc-pV5Z electronic basis set. The carbon atom is positioned at -1.058~Å, the nitrogen atom is positioned at -2.206~Å. and the proton basis functions are centered at the origin. The on-axis proton density is along the line that connects the carbon and nitrogen atoms.}
  \label{fig:hcn_density}
\end{figure}

\begin{table}[]
\caption{Root-Mean-Square deviation (RMSD)\footnote{The RMSD values are given in atomic units. The RMSD is calculated as the square root of the average of the squares of the density differences between the proton densities obtained with the NEO and the FGH methods for every grid point.} values of the proton density calculated with a NEO method relative to the FGH reference density.\footnote{The FGH reference density was obtained at the conventional CCSD/aug-cc-pVDZ level of theory.}}
\begin{tabular}{|c|cccc|cccc|}\hline\hline
 & \multicolumn{4}{c|}{FHF$^-$\footnote{The fluorine atoms are positioned at -1.1335~Å and 1.1335~Å. The cubic grid with 32 points in each direction spans the range from -0.5610~Å to 0.5984~Å.}} & \multicolumn{4}{c|}{HCN\footnote{The carbon atom is positioned at -1.058~Å and the nitrogen atom is positioned at -2.206~Å. The cubic grid with 32 points in each direction spans the range from -0.7258~Å to 0.7742~Å.}} \\\hline
method & \multicolumn{1}{c|}{aDZ} & \multicolumn{1}{c|}{aTZ} & \multicolumn{1}{c|}{aQZ} & a5Z & \multicolumn{1}{c|}{aDZ} & \multicolumn{1}{c|}{aTZ} & \multicolumn{1}{c|}{aQZ} & a5Z \\\hline
NEO-HF & \multicolumn{1}{c|}{0.75} & \multicolumn{1}{c|}{0.75} & \multicolumn{1}{c|}{0.73} & 0.73 & \multicolumn{1}{c|}{0.75} & \multicolumn{1}{c|}{0.75} & \multicolumn{1}{c|}{0.74} & 0.73 \\
NEO-CCSD(ep) & \multicolumn{1}{c|}{0.41} & \multicolumn{1}{c|}{0.26} & \multicolumn{1}{c|}{0.15} & 0.13 & \multicolumn{1}{c|}{0.48} & \multicolumn{1}{c|}{0.36} & \multicolumn{1}{c|}{0.25} & 0.23 \\
NEO-CCSD(eep) & \multicolumn{1}{c|}{0.39} & \multicolumn{1}{c|}{0.21} & \multicolumn{1}{c|}{0.08} & 0.06 & \multicolumn{1}{c|}{0.46} & \multicolumn{1}{c|}{0.32} & \multicolumn{1}{c|}{0.20} & 0.18 \\
NEO-CC2 & \multicolumn{1}{c|}{0.44} & \multicolumn{1}{c|}{0.33} & \multicolumn{1}{c|}{0.25} & 0.23 & \multicolumn{1}{c|}{0.50} & \multicolumn{1}{c|}{0.43} & \multicolumn{1}{c|}{0.36} & 0.34\\
NEO-SOS-CC2 & \multicolumn{1}{c|}{0.45} & \multicolumn{1}{c|}{0.34} & \multicolumn{1}{c|}{0.26} & 0.24 & \multicolumn{1}{c|}{0.51} & \multicolumn{1}{c|}{0.44} & \multicolumn{1}{c|}{0.37} & 0.35\\
NEO-SOS$'$-CC2 & \multicolumn{1}{c|}{-} & \multicolumn{1}{c|}{-} & \multicolumn{1}{c|}{0.06} & - & \multicolumn{1}{c|}{-} & \multicolumn{1}{c|}{-} & \multicolumn{1}{c|}{0.18} & -\\\hline\hline
\end{tabular}
\label{table:RMSD}
\end{table}

To assess the accuracy of the novel NEO methods, we computed the proton densities for the FHF$^-$ molecule and the HCN molecule. The benchmark densities were calculated with the Fourier Grid Hamiltonian (FGH) method,~\cite{webb2000fourier} which is numerically nearly exact for these two systems. The FGH reference density was obtained at the conventional CCSD/aug-cc-pVDZ level of theory. For consistency with these NEO calculations, in the FGH method, only the hydrogen was treated quantum mechanically and the other nuclei were fixed. Figures~\ref{fig:fhf_density} and~\ref{fig:hcn_density} show on-axis one-dimensional slices of the proton densities for these two molecules calculated with the NEO-HF, NEO-CC2, NEO-CCSD(ep), NEO-CCSD(eep), and FGH methods. The on-axis proton density is along the line that connects the heavy atoms, which are either two fluorine atoms (for FHF$^-$) or carbon and nitrogen atoms (for HCN). Both the NEO and FGH three-dimensional proton densities are normalized to unity. 

To quantify the difference between the proton densities obtained with the NEO methods and the FGH method, we computed the root-mean-square deviations (RMSDs). The RMSD values for both the FHF$^-$ and HCN molecules calculated with the NEO-HF, NEO-CCSD(ep), NEO-CCSD(eep), NEO-CC2, and NEO-SOS$'$-CC2 methods and with different basis sets are given in Table~\ref{table:RMSD}. As discussed previously,\cite{pavosevic2018ccsd,pavosevic2020chemrev} the NEO-HF method produces proton densities that are too localized, mainly due to the inadequacies of the mean-field description. This behavior is depicted by the solid red curves in Figs.~\ref{fig:fhf_density} and~\ref{fig:hcn_density}. Additionally, Table~\ref{table:RMSD} shows that the NEO-HF exhibits the largest RMSD values, which remain nearly constant with increase of the basis set size. 

\begin{table*}[]
\caption{Absolute Deviation and Mean Unsigned Error (MUE) of Proton Affinities
(in electron volts) with Respect to Experimental Data.\footnote{Experimental values obtained from Refs.~\citenum{cumming1978summary,graul1990gas,hunter1998evaluated}}}
\begin{tabular}{|c|c|cccC{0.95cm}|cccC{0.95cm}|cccC{0.95cm}|c|}
\hline\hline
 &  & \multicolumn{4}{c|}{NEO-CCSD(ep)} & \multicolumn{4}{c|}{NEO-CCSD(eep)} &
 \multicolumn{4}{c|}{NEO-CC2} & \multicolumn{1}{C{1.5cm}|}{NEO-SOS$'$-CC2}    \\\hline
molecule & experiment & \multicolumn{1}{C{0.95cm}}{aDZ} & \multicolumn{1}{C{0.95cm}}{aTZ} & \multicolumn{1}{C{0.95cm}}{aQZ} & a5Z & \multicolumn{1}{C{0.95cm}}{aDZ} & \multicolumn{1}{C{0.95cm}}{aTZ} & \multicolumn{1}{C{0.95cm}}{aQZ} & a5Z &
\multicolumn{1}{C{0.95cm}}{aDZ} & \multicolumn{1}{C{0.95cm}}{aTZ} & \multicolumn{1}{C{0.95cm}}{aQZ} & a5Z &\multicolumn{1}{c|}{aQZ} \\\hline
CN$^-$ & 15.31 & \multicolumn{1}{c}{0.58} & \multicolumn{1}{c}{0.30} & \multicolumn{1}{c}{0.24} & 0.18 & \multicolumn{1}{c}{0.55} & \multicolumn{1}{c}{0.26} & \multicolumn{1}{c}{0.18} & 0.12 &
\multicolumn{1}{c}{0.71} & \multicolumn{1}{c}{0.46} & \multicolumn{1}{c}{0.43} & 0.38 &\multicolumn{1}{c|}{0.01} \\
NO$_2^-$ & 14.75 & \multicolumn{1}{c}{0.40} & \multicolumn{1}{c}{0.20} & \multicolumn{1}{c}{0.13} & 0.08 & \multicolumn{1}{c}{0.36} & \multicolumn{1}{c}{0.15} & \multicolumn{1}{c}{0.06} & $<$0.01 &
\multicolumn{1}{c}{0.76} & \multicolumn{1}{c}{0.59} & \multicolumn{1}{c}{0.53} & 0.49 &\multicolumn{1}{c|}{0.03} \\
NH$_3$ & 8.85 & \multicolumn{1}{c}{0.35} & \multicolumn{1}{c}{0.20} & \multicolumn{1}{c}{0.11} & 0.05 & \multicolumn{1}{c}{0.32} & \multicolumn{1}{c}{0.16} & \multicolumn{1}{c}{0.06} & $<$0.01 &
\multicolumn{1}{c}{0.47} & \multicolumn{1}{c}{0.35} & \multicolumn{1}{c}{0.29} & 0.24 &\multicolumn{1}{c|}{0.08} \\
HCOO$^-$ & 14.97 & \multicolumn{1}{c}{0.42} & \multicolumn{1}{c}{0.22} & \multicolumn{1}{c}{0.12} & 0.05 & \multicolumn{1}{c}{0.39} & \multicolumn{1}{c}{0.18} & \multicolumn{1}{c}{0.07} & 0.01 &
\multicolumn{1}{c}{0.74} & \multicolumn{1}{c}{0.58} & \multicolumn{1}{c}{0.51} & 0.43 &\multicolumn{1}{c|}{0.02} \\
HO$^-$ & 16.95 & \multicolumn{1}{c}{0.46} & \multicolumn{1}{c}{0.21} & \multicolumn{1}{c}{0.12} & 0.05 & \multicolumn{1}{c}{0.42} & \multicolumn{1}{c}{0.16} & \multicolumn{1}{c}{0.05} & 0.02 &
\multicolumn{1}{c}{0.83} & \multicolumn{1}{c}{0.64} & \multicolumn{1}{c}{0.58} & 0.54 &\multicolumn{1}{c|}{0.11} \\
HS$^-$ & 15.31 & \multicolumn{1}{c}{0.55} & \multicolumn{1}{c}{0.32} & \multicolumn{1}{c}{0.25} & 0.14 & \multicolumn{1}{c}{0.52} & \multicolumn{1}{c}{0.27} & \multicolumn{1}{c}{0.18} & 0.07 &
\multicolumn{1}{c}{0.74} & \multicolumn{1}{c}{0.51} & \multicolumn{1}{c}{0.47} & 0.38 &\multicolumn{1}{c|}{0.04} \\
H$_2$O & 7.16 & \multicolumn{1}{c}{0.42} & \multicolumn{1}{c}{0.23} & \multicolumn{1}{c}{0.16} & 0.10 & \multicolumn{1}{c}{0.39} & \multicolumn{1}{c}{0.19} & \multicolumn{1}{c}{0.10} & 0.05 &
\multicolumn{1}{c}{0.55} & \multicolumn{1}{c}{0.39} & \multicolumn{1}{c}{0.33} & 0.29 &\multicolumn{1}{c|}{$<$0.01} \\
H$_2$S & 7.31 & \multicolumn{1}{c}{0.34} & \multicolumn{1}{c}{0.20} & \multicolumn{1}{c}{0.13} & 0.04 & \multicolumn{1}{c}{0.31} & \multicolumn{1}{c}{0.15} & \multicolumn{1}{c}{0.07} & 0.02 &
\multicolumn{1}{c}{0.47} & \multicolumn{1}{c}{0.29} & \multicolumn{1}{c}{0.17} & 0.26 &\multicolumn{1}{c|}{0.11} \\
CO & 6.16 & \multicolumn{1}{c}{0.40} & \multicolumn{1}{c}{0.21} & \multicolumn{1}{c}{0.16} & 0.11 & \multicolumn{1}{c}{0.38} & \multicolumn{1}{c}{0.17} & \multicolumn{1}{c}{0.11} & 0.06 &
\multicolumn{1}{c}{0.36} & \multicolumn{1}{c}{0.19} & \multicolumn{1}{c}{0.17} & 0.13 &\multicolumn{1}{c|}{0.12} \\
N$_2$ & 5.12 & \multicolumn{1}{c}{0.43} & \multicolumn{1}{c}{0.24} & \multicolumn{1}{c}{0.17} & 0.14 & \multicolumn{1}{c}{0.41} & \multicolumn{1}{c}{0.20} & \multicolumn{1}{c}{0.12} & 0.08 &
\multicolumn{1}{c}{0.52} & \multicolumn{1}{c}{0.35} & \multicolumn{1}{c}{0.30} & 0.28 &\multicolumn{1}{c|}{0.04} \\
CO$_2$ & 5.60 & \multicolumn{1}{c}{0.37} & \multicolumn{1}{c}{0.22} & \multicolumn{1}{c}{0.17} & 0.12 & \multicolumn{1}{c}{0.34} & \multicolumn{1}{c}{0.18} & \multicolumn{1}{c}{0.11} & 0.06 &
\multicolumn{1}{c}{0.57} & \multicolumn{1}{c}{0.44} & \multicolumn{1}{c}{0.40} & 0.36 &\multicolumn{1}{c|}{0.01} \\
CH$_2$O & 7.39 & \multicolumn{1}{c}{0.35} & \multicolumn{1}{c}{0.19} & \multicolumn{1}{c}{0.11} & 0.06 & \multicolumn{1}{c}{0.31} & \multicolumn{1}{c}{0.14} & \multicolumn{1}{c}{0.05} & $<$0.01 &
\multicolumn{1}{c}{0.52} & \multicolumn{1}{c}{0.40} & \multicolumn{1}{c}{0.34} & 0.30 &\multicolumn{1}{c|}{0.04} \\\hline
MUE &  & \multicolumn{1}{c}{0.42} & \multicolumn{1}{c}{0.23} & \multicolumn{1}{c}{0.16} & 0.09 & \multicolumn{1}{c}{0.39} & \multicolumn{1}{c}{0.18} & \multicolumn{1}{c}{0.09} & 0.04 &
\multicolumn{1}{c}{0.60} & \multicolumn{1}{c}{0.44} & \multicolumn{1}{c}{0.39} & 0.34 &\multicolumn{1}{c|}{0.05} \\\hline\hline

\end{tabular}

\label{table:PA}
\end{table*}

Inclusion of the correlation effects between quantum particles with any of the studied NEO-CC methods significantly improves the calculated proton densities. Because the NEO-CC2, NEO-CCSD(ep), and NEO-CCSD(eep) methods constitute a NEO-CC hierarchy, the calculated proton densities are improved in that order, where NEO-CCSD(eep) exhibits the highest degree of accuracy, as depicted in Figs.~\ref{fig:fhf_density} and~\ref{fig:hcn_density} for both systems.  Moreover, Table~\ref{table:RMSD} also indicates that the NEO-CC2 method yields the largest RMSD, whereas the NEO-CCSD(eep) method yields the lowest RMSD among the studied NEO-CC methods. Table~\ref{table:RMSD} also shows that the proton density depends strongly on the electronic basis set size for all of these NEO-CC methods and that very extensive electronic basis sets are required for achieving quantitative accuracy. The performance of the NEO-CCSD(ep) method is closer to that of the NEO-CCSD(eep) method than to the NEO-CC2 method for the proton densities. Nevertheless, inclusion of the triple electron-electron-proton excitations significantly improves upon the  NEO-CCSD(ep) method. Thus, inclusion of even higher order excitations in the NEO-CC methods is expected to provide more accurate proton densities that approach the FGH reference curve in a systematic manner but at higher computational cost. 

As introduced previously in the context of the NEO-OOMP2 method,~\cite{pavosevic2020multicomponent} the performance of the NEO-CC2 method can be enhanced by the SOS$'$ approach, in which different correlation energy contributions are scaled with additional parameters. For the NEO-SOS-CC2 method, the different spin contributions to the electron-electron correlation energy are scaled with parameters $c_{\text{os}} = 1.3$ and $c_{\text{ss}} = 0.0$ for the opposite-spin and same-spin electron-electron correlation contributions, respectively. These parameters were determined previously in the context of electronic structure methods.\cite{jung2004scaled} As shown in Table~\ref{table:RMSD}, this parametrization for NEO-SOS-CC2 gives almost the same proton density RMSDs as those obtained with the NEO-CC2 method, mainly because the protonic orbitals are not influenced significantly when only the electron correlation energy is modified. In the NEO-SOS$'$-CC2 method, the electron-proton correlation energy is also scaled by the parameter  $c_{\text{ep}}$. We determined the optimal $c_{\text{ep}} = 1.6$ parameter by minimizing the difference between the proton density RMSD calculated with the  NEO-SOS$'$-CC2/aug-cc-pVQZ and NEO-CCSD(eep)/aug-cc-pV5Z methods. The proton densities calculated with the NEO-SOS$'$-CC2 method for both molecular systems are virtually
indistinguishable from those calculated with the NEO-CCSD(eep) method, and therefore they are not included in Figs.~\ref{fig:fhf_density} and~\ref{fig:hcn_density} for clarity.

To further test the accuracy of these NEO methods, we used them to calculate the proton affinities for a set of 12 small molecules~\cite{pavosevic2018ccsd} and compared the predicted proton affinities to the experimental values. Within the NEO framework, the proton affinity of a species A is calculated as  PA(A) = $E_{\text{A}}-E_{\text{HA}^+}+5/2RT$.~\cite{diaz2013generalized,brorsen2017multicomponent} In this expression, $E_{\text{A}}$ is the energy of species A calculated with the conventional electronic structure method, and $E_{\text{HA}^+}$ is the energy of species HA$^+$ calculated with the corresponding NEO method, where the hydrogen H is treated quantum mechanically. The last term in this expression, $5/2RT$ ($R$ and $T$ are the ideal gas constant and the temperature, respectively), accounts for conversion from energy to enthalpy and the change in translational energy. Because the NEO method inherently includes the zero-point energy contribution of the quantum proton into the energy calculation, this approach does not require calculation of a Hessian. The vibrational zero-point energies associated with the other nuclei are assumed to be unchanged upon protonation, which has been shown to be a reasonable approximation.~\cite{pavosevic2018ccsd} 

Table~\ref{table:PA} presents the absolute deviations of the calculated proton affinities from experimental data, as obtained with different NEO methods and electronic basis sets. The results indicate a systematic improvement of the calculated proton affinities as the basis set cardinal number (X=D,T,Q,5) increases. Because the NEO-CCSD(eep) recovers a greater amount of the correlation energy, the results obtained with this method are more accurate than the results obtained with the NEO-CCSD(ep) method. For a large electronic basis set, aug-cc-pV5Z, the NEO-CCSD(eep) method provides MUEs that are within both chemical ($\sim0.05$ eV) and experimental ($\sim0.09$ eV) accuracy.~\cite{hunter1998evaluated} Note that the NEO-CCSD(ep) method does not produce an MUE that is within chemical accuracy even with the largest electronic basis set employed.  These results show that in order to achieve chemical accuracy with the NEO-CC methods, it is important to incorporate the triple excitations where two electrons and one proton are excited simultaneously. In our previous work,~\cite{pavosevic2018ccsd} the missing electron-proton correlation of the NEO-CCSD(ep) method due to these triple excitations was included by using a larger electronic basis set for the hydrogen nuclei treated quantum mechanically than for the other nuclei. In particular, the calculations were performed with the aug-cc-pVTZ basis set for the classical nuclei and the aug-cc-pVQZ basis set for the quantum nuclei. This combination of electronic basis sets produced an MUE of 0.04~eV for these proton affinities with the NEO-CCSD(ep) method, as reported previously in Ref.~\citenum{pavosevic2018ccsd}. With the NEO-CCSD(eep),  chemical accuracy is achieved without such mixed basis sets.

In the future, the basis set incompleteness error could be tackled by a NEO variant of explicitly correlated methods.\cite{explicitly} Alternatively, the basis set incompleteness error can be mitigated by a basis set extrapolation scheme.\cite{extrapolation} Although the different correlation energy contributions (i.e., electron-electron and electron-proton) have a different rate of convergence to the complete basis set, here we used the established extrapolation scheme~\cite{extrapolation} developed for electron correlation and applied it to the correlation energy obtained with the NEO-CCSD methods with the aug-cc-pVTZ and aug-cc-pVQZ basis sets. The resulting MUEs for the NEO-CCSD(ep) and NEO-CCSD(eep) methods are 0.10~eV and 0.05~eV, respectively, which are in excellent agreement with the results obtained with the aug-cc-pV5Z basis set (0.09~eV and 0.04~eV, respectively). Developing strategies to handle the slow convergence of the basis set represents an interesting research direction for the future. 

The NEO-CC2 method provides the largest MUE among the NEO-CC methods due to the inadequate treatment of the correlations between quantum particles. Scaling of the electron-electron and electron-proton correlation contributions to the correlation energy in the NEO-SOS$'$-CC2 method reduces the MUE to only 0.05~eV, which is in excellent agreement with the NEO-CCSD(eep)/aug-cc-pV5Z method (MUE of 0.04~eV). The employed scaling parameters are $c_{\text{os}}=1.3$, $c_{\text{ss}}=0.0$, and $c_{\text{ep}}=1.6$, as determined above by fitting to the proton densities. Interestingly, we found that this value of $c_{\text{ep}}$ is also optimal for proton affinities. Therefore, the NEO-SOS$'$-CC2 method provides results that are within both chemical ($\sim0.05$eV) and experimental ($\sim0.09$eV) accuracy, making it a viable computationally efficient alternative to the NEO-CCSD(eep) method.

\section{Conclusions}
This paper presents the NEO-CCSD(eep) method, which includes simultaneous double electronic excitations and single protonic excitations.  The proton densities of the FHF$^-$ and HCN molecules computed with this method are in excellent agreement with the grid-based reference, outperforming all previously studied NEO methods. The $\Lambda$-equations that are necessary for obtaining the protonic density are calculated with automatic differentiation, which does not require explicit implementation of these equations. 
Our calculations also illustrate that the NEO-CCSD(eep) method, in conjunction with consistent basis sets, produces proton affinities within experimental and chemical accuracy, in contrast to lower-level NEO-CC methods. These results demonstrate the importance of the triple electron-electron-proton excitations for a quantitatively accurate description of the nuclear quantum effects. 

We also developed and tested the NEO-CC2 method. As a rather crude approximation of NEO-CCSD(ep) and NEO-CCSD(eep), the properties predicted with NEO-CC2 are not accurate. The related NEO-SOS$'$-CC2 method, which scales the same-spin and opposite-spin components of the electron-electron correlation energy and the electron-proton correlation energy, achieves nearly the same level of accuracy as the NEO-CCSD(eep) method. An appealing feature of the NEO-SOS$'$-CC2 method is that it can be implemented with $\mathcal{O}(N^4)$ computational scaling, and it can be used as an alternative to the NEO-CCSD(eep) method for large molecular systems. 

Moreover, the NEO-SOS$'$-CC2 method can be extended to treat excited states. A key advantage of this method is that it will be suitable for the description of excitations with double excitation character, in which both an electron and a proton are excited simultaneously. Such excitations correspond to an excited proton vibrational state associated with an excited electronic state, and they are essential for various photochemical processes such as photoinduced proton transfer and proton-coupled electron transfer. Lastly, this work shows that the NEO-CCSD(eep) method can serve as a reference in parametrization of the computationally more efficient methods, such as NEO-DFT and NEO-SOS$'$-MP2, as it was used directly here for parametrizing the NEO-SOS$'$-CC2 method. Thus, the developments presented in this work open up many research paths for future theoretical
developments and applications to systems exhibiting significant nuclear quantum effects.


\begin{acknowledgments}
The authors thank Dr. Jonathan Fetherolf for helpful discussions and Dr. Kurt Brorsen for useful discussions about electronic basis sets. This work was supported in part by the National Science Foundation Grant No. CHE-1954348 (S.H.-S.). The Flatiron Institute is a division of the Simons Foundation.
\end{acknowledgments}

\section*{AUTHOR DECLARATIONS}

\textbf{Conflict of Interest}\\ 

The authors have no conflicts of interest to disclose.

\section*{Data Availability Statement}

The data that support the findings of this study are available within this article.

\bibliography{Journal_Short_Name,references}

\begin{thebibliography}{54}%
\makeatletter
\providecommand \@ifxundefined [1]{%
 \@ifx{#1\undefined}
}%
\providecommand \@ifnum [1]{%
 \ifnum #1\expandafter \@firstoftwo
 \else \expandafter \@secondoftwo
 \fi
}%
\providecommand \@ifx [1]{%
 \ifx #1\expandafter \@firstoftwo
 \else \expandafter \@secondoftwo
 \fi
}%
\providecommand \natexlab [1]{#1}%
\providecommand \enquote  [1]{``#1''}%
\providecommand \bibnamefont  [1]{#1}%
\providecommand \bibfnamefont [1]{#1}%
\providecommand \citenamefont [1]{#1}%
\providecommand \href@noop [0]{\@secondoftwo}%
\providecommand \href [0]{\begingroup \@sanitize@url \@href}%
\providecommand \@href[1]{\@@startlink{#1}\@@href}%
\providecommand \@@href[1]{\endgroup#1\@@endlink}%
\providecommand \@sanitize@url [0]{\catcode `\\12\catcode `\$12\catcode
  `\&12\catcode `\#12\catcode `\^12\catcode `\_12\catcode `\%12\relax}%
\providecommand \@@startlink[1]{}%
\providecommand \@@endlink[0]{}%
\providecommand \url  [0]{\begingroup\@sanitize@url \@url }%
\providecommand \@url [1]{\endgroup\@href {#1}{\urlprefix }}%
\providecommand \urlprefix  [0]{URL }%
\providecommand \Eprint [0]{\href }%
\providecommand \doibase [0]{http://dx.doi.org/}%
\providecommand \selectlanguage [0]{\@gobble}%
\providecommand \bibinfo  [0]{\@secondoftwo}%
\providecommand \bibfield  [0]{\@secondoftwo}%
\providecommand \translation [1]{[#1]}%
\providecommand \BibitemOpen [0]{}%
\providecommand \bibitemStop [0]{}%
\providecommand \bibitemNoStop [0]{.\EOS\space}%
\providecommand \EOS [0]{\spacefactor3000\relax}%
\providecommand \BibitemShut  [1]{\csname bibitem#1\endcsname}%
\let\auto@bib@innerbib\@empty
\bibitem [{\citenamefont {Pavosevic}, \citenamefont {Culpitt},\ and\
  \citenamefont {Hammes-Schiffer}(2020)}]{pavosevic2020chemrev}%
  \BibitemOpen
  \bibfield  {author} {\bibinfo {author} {\bibfnamefont {F.}~\bibnamefont
  {Pavosevic}}, \bibinfo {author} {\bibfnamefont {T.}~\bibnamefont {Culpitt}},
  \ and\ \bibinfo {author} {\bibfnamefont {S.}~\bibnamefont
  {Hammes-Schiffer}},\ }\bibfield  {title} {\enquote {\bibinfo {title}
  {Multicomponent quantum chemistry: Integrating electronic and nuclear quantum
  effects via the nuclear--electronic orbital method},}\ }\href@noop {}
  {\bibfield  {journal} {\bibinfo  {journal} {Chem. Rev.}\ }\textbf {\bibinfo
  {volume} {120}},\ \bibinfo {pages} {4222--4253} (\bibinfo {year}
  {2020})}\BibitemShut {NoStop}%
\bibitem [{\citenamefont {Ruggenthaler}\ \emph {et~al.}(2018)\citenamefont
  {Ruggenthaler}, \citenamefont {Tancogne-Dejean}, \citenamefont {Flick},
  \citenamefont {Appel},\ and\ \citenamefont
  {Rubio}}]{ruggenthaler2018quantum}%
  \BibitemOpen
  \bibfield  {author} {\bibinfo {author} {\bibfnamefont {M.}~\bibnamefont
  {Ruggenthaler}}, \bibinfo {author} {\bibfnamefont {N.}~\bibnamefont
  {Tancogne-Dejean}}, \bibinfo {author} {\bibfnamefont {J.}~\bibnamefont
  {Flick}}, \bibinfo {author} {\bibfnamefont {H.}~\bibnamefont {Appel}}, \ and\
  \bibinfo {author} {\bibfnamefont {A.}~\bibnamefont {Rubio}},\ }\bibfield
  {title} {\enquote {\bibinfo {title} {From a quantum-electrodynamical
  light-matter description to novel spectroscopies},}\ }\href@noop {}
  {\bibfield  {journal} {\bibinfo  {journal} {Nat. Rev. Chem.}\ }\textbf
  {\bibinfo {volume} {2}},\ \bibinfo {pages} {1--16} (\bibinfo {year}
  {2018})}\BibitemShut {NoStop}%
\bibitem [{\citenamefont {Pavo{\v{s}}evi{\'c}}\ \emph
  {et~al.}(2022)\citenamefont {Pavo{\v{s}}evi{\'c}}, \citenamefont
  {Hammes-Schiffer}, \citenamefont {Rubio},\ and\ \citenamefont
  {Flick}}]{pavosevic2021cavity}%
  \BibitemOpen
  \bibfield  {author} {\bibinfo {author} {\bibfnamefont {F.}~\bibnamefont
  {Pavo{\v{s}}evi{\'c}}}, \bibinfo {author} {\bibfnamefont {S.}~\bibnamefont
  {Hammes-Schiffer}}, \bibinfo {author} {\bibfnamefont {A.}~\bibnamefont
  {Rubio}}, \ and\ \bibinfo {author} {\bibfnamefont {J.}~\bibnamefont
  {Flick}},\ }\bibfield  {title} {\enquote {\bibinfo {title} {Cavity-modulated
  proton transfer reactions},}\ }\href@noop {} {\bibfield  {journal} {\bibinfo
  {journal} {J. Am. Chem. Soc.}\ }\textbf {\bibinfo {volume} {144}},\ \bibinfo
  {pages} {4995–5002} (\bibinfo {year} {2022})}\BibitemShut {NoStop}%
\bibitem [{\citenamefont {Ishimoto}, \citenamefont {Tachikawa},\ and\
  \citenamefont {Nagashima}(2009)}]{ishimoto2009review}%
  \BibitemOpen
  \bibfield  {author} {\bibinfo {author} {\bibfnamefont {T.}~\bibnamefont
  {Ishimoto}}, \bibinfo {author} {\bibfnamefont {M.}~\bibnamefont {Tachikawa}},
  \ and\ \bibinfo {author} {\bibfnamefont {U.}~\bibnamefont {Nagashima}},\
  }\bibfield  {title} {\enquote {\bibinfo {title} {Review of multicomponent
  molecular orbital method for direct treatment of nuclear quantum effect},}\
  }\href@noop {} {\bibfield  {journal} {\bibinfo  {journal} {Int. J. Quant.
  Chem.}\ }\textbf {\bibinfo {volume} {109}},\ \bibinfo {pages} {2677--2694}
  (\bibinfo {year} {2009})}\BibitemShut {NoStop}%
\bibitem [{\citenamefont {Reyes}, \citenamefont {Moncada},\ and\ \citenamefont
  {Charry}(2019)}]{reyes2019any}%
  \BibitemOpen
  \bibfield  {author} {\bibinfo {author} {\bibfnamefont {A.}~\bibnamefont
  {Reyes}}, \bibinfo {author} {\bibfnamefont {F.}~\bibnamefont {Moncada}}, \
  and\ \bibinfo {author} {\bibfnamefont {J.}~\bibnamefont {Charry}},\
  }\bibfield  {title} {\enquote {\bibinfo {title} {The any particle molecular
  orbital approach: A short review of the theory and applications},}\
  }\href@noop {} {\bibfield  {journal} {\bibinfo  {journal} {Int. J. Quant.
  Chem.}\ }\textbf {\bibinfo {volume} {119}},\ \bibinfo {pages} {e25705}
  (\bibinfo {year} {2019})}\BibitemShut {NoStop}%
\bibitem [{\citenamefont {Webb}, \citenamefont {Iordanov},\ and\ \citenamefont
  {Hammes-Schiffer}(2002)}]{webb2002multiconfigurational}%
  \BibitemOpen
  \bibfield  {author} {\bibinfo {author} {\bibfnamefont {S.~P.}\ \bibnamefont
  {Webb}}, \bibinfo {author} {\bibfnamefont {T.}~\bibnamefont {Iordanov}}, \
  and\ \bibinfo {author} {\bibfnamefont {S.}~\bibnamefont {Hammes-Schiffer}},\
  }\bibfield  {title} {\enquote {\bibinfo {title} {Multiconfigurational
  nuclear-electronic orbital approach: Incorporation of nuclear quantum effects
  in electronic structure calculations},}\ }\href@noop {} {\bibfield  {journal}
  {\bibinfo  {journal} {J. Chem. Phys.}\ }\textbf {\bibinfo {volume} {117}},\
  \bibinfo {pages} {4106--4118} (\bibinfo {year} {2002})}\BibitemShut {NoStop}%
\bibitem [{\citenamefont {Pavošević}, \citenamefont {Culpitt},\ and\
  \citenamefont {Hammes-Schiffer}(2018)}]{pavosevic2018ccsd}%
  \BibitemOpen
  \bibfield  {author} {\bibinfo {author} {\bibfnamefont {F.}~\bibnamefont
  {Pavošević}}, \bibinfo {author} {\bibfnamefont {T.}~\bibnamefont
  {Culpitt}}, \ and\ \bibinfo {author} {\bibfnamefont {S.}~\bibnamefont
  {Hammes-Schiffer}},\ }\bibfield  {title} {\enquote {\bibinfo {title}
  {Multicomponent coupled cluster singles and doubles theory within the
  nuclear-electronic orbital framework},}\ }\href@noop {} {\bibfield  {journal}
  {\bibinfo  {journal} {J. Chem. Theory Comput.}\ }\textbf {\bibinfo {volume}
  {15}},\ \bibinfo {pages} {338--347} (\bibinfo {year} {2018})}\BibitemShut
  {NoStop}%
\bibitem [{\citenamefont {Pavo{\v{s}}evi{\'c}}\ and\ \citenamefont
  {Hammes-Schiffer}(2019{\natexlab{a}})}]{pavovsevic2019multicomponent}%
  \BibitemOpen
  \bibfield  {author} {\bibinfo {author} {\bibfnamefont {F.}~\bibnamefont
  {Pavo{\v{s}}evi{\'c}}}\ and\ \bibinfo {author} {\bibfnamefont
  {S.}~\bibnamefont {Hammes-Schiffer}},\ }\bibfield  {title} {\enquote
  {\bibinfo {title} {Multicomponent coupled cluster singles and doubles and
  brueckner doubles methods: Proton densities and energies},}\ }\href@noop {}
  {\bibfield  {journal} {\bibinfo  {journal} {J. Chem. Phys.}\ }\textbf
  {\bibinfo {volume} {151}},\ \bibinfo {pages} {074104} (\bibinfo {year}
  {2019}{\natexlab{a}})}\BibitemShut {NoStop}%
\bibitem [{\citenamefont {Pak}, \citenamefont {Chakraborty},\ and\
  \citenamefont {Hammes-Schiffer}(2007)}]{pak2007density}%
  \BibitemOpen
  \bibfield  {author} {\bibinfo {author} {\bibfnamefont {M.~V.}\ \bibnamefont
  {Pak}}, \bibinfo {author} {\bibfnamefont {A.}~\bibnamefont {Chakraborty}}, \
  and\ \bibinfo {author} {\bibfnamefont {S.}~\bibnamefont {Hammes-Schiffer}},\
  }\bibfield  {title} {\enquote {\bibinfo {title} {Density functional theory
  treatment of electron correlation in the nuclear- electronic orbital
  approach},}\ }\href@noop {} {\bibfield  {journal} {\bibinfo  {journal} {The
  Journal of Physical Chemistry A}\ }\textbf {\bibinfo {volume} {111}},\
  \bibinfo {pages} {4522--4526} (\bibinfo {year} {2007})}\BibitemShut {NoStop}%
\bibitem [{\citenamefont {Yang}\ \emph {et~al.}(2017)\citenamefont {Yang},
  \citenamefont {Brorsen}, \citenamefont {Culpitt}, \citenamefont {Pak},\ and\
  \citenamefont {Hammes-Schiffer}}]{yang2017development}%
  \BibitemOpen
  \bibfield  {author} {\bibinfo {author} {\bibfnamefont {Y.}~\bibnamefont
  {Yang}}, \bibinfo {author} {\bibfnamefont {K.~R.}\ \bibnamefont {Brorsen}},
  \bibinfo {author} {\bibfnamefont {T.}~\bibnamefont {Culpitt}}, \bibinfo
  {author} {\bibfnamefont {M.~V.}\ \bibnamefont {Pak}}, \ and\ \bibinfo
  {author} {\bibfnamefont {S.}~\bibnamefont {Hammes-Schiffer}},\ }\bibfield
  {title} {\enquote {\bibinfo {title} {Development of a practical
  multicomponent density functional for electron-proton correlation to produce
  accurate proton densities},}\ }\href@noop {} {\bibfield  {journal} {\bibinfo
  {journal} {J. Chem. Phys.}\ }\textbf {\bibinfo {volume} {147}},\ \bibinfo
  {pages} {114113} (\bibinfo {year} {2017})}\BibitemShut {NoStop}%
\bibitem [{\citenamefont {Brorsen}, \citenamefont {Yang},\ and\ \citenamefont
  {Hammes-Schiffer}(2017)}]{brorsen2017multicomponent}%
  \BibitemOpen
  \bibfield  {author} {\bibinfo {author} {\bibfnamefont {K.~R.}\ \bibnamefont
  {Brorsen}}, \bibinfo {author} {\bibfnamefont {Y.}~\bibnamefont {Yang}}, \
  and\ \bibinfo {author} {\bibfnamefont {S.}~\bibnamefont {Hammes-Schiffer}},\
  }\bibfield  {title} {\enquote {\bibinfo {title} {Multicomponent density
  functional theory: Impact of nuclear quantum effects on proton affinities and
  geometries},}\ }\href@noop {} {\bibfield  {journal} {\bibinfo  {journal} {J.
  Phys. Chem. Lett.}\ }\textbf {\bibinfo {volume} {8}},\ \bibinfo {pages}
  {3488--3493} (\bibinfo {year} {2017})}\BibitemShut {NoStop}%
\bibitem [{\citenamefont {Nakai}\ and\ \citenamefont
  {Sodeyama}(2003)}]{nakai2003many}%
  \BibitemOpen
  \bibfield  {author} {\bibinfo {author} {\bibfnamefont {H.}~\bibnamefont
  {Nakai}}\ and\ \bibinfo {author} {\bibfnamefont {K.}~\bibnamefont
  {Sodeyama}},\ }\bibfield  {title} {\enquote {\bibinfo {title} {Many-body
  effects in nonadiabatic molecular theory for simultaneous determination of
  nuclear and electronic wave functions: Ab initio nomo/mbpt and cc methods},}\
  }\href@noop {} {\bibfield  {journal} {\bibinfo  {journal} {The Journal of
  chemical physics}\ }\textbf {\bibinfo {volume} {118}},\ \bibinfo {pages}
  {1119--1127} (\bibinfo {year} {2003})}\BibitemShut {NoStop}%
\bibitem [{\citenamefont {Pavošević}, \citenamefont {Rousseau},\ and\
  \citenamefont {Hammes-Schiffer}(2020)}]{pavosevic2020multicomponent}%
  \BibitemOpen
  \bibfield  {author} {\bibinfo {author} {\bibfnamefont {F.}~\bibnamefont
  {Pavošević}}, \bibinfo {author} {\bibfnamefont {B.~J.}\ \bibnamefont
  {Rousseau}}, \ and\ \bibinfo {author} {\bibfnamefont {S.}~\bibnamefont
  {Hammes-Schiffer}},\ }\bibfield  {title} {\enquote {\bibinfo {title}
  {Multicomponent orbital-optimized perturbation theory methods: Approaching
  coupled cluster accuracy at lower cost},}\ }\href@noop {} {\bibfield
  {journal} {\bibinfo  {journal} {J. Phys. Chem. Lett.}\ }\textbf {\bibinfo
  {volume} {11}},\ \bibinfo {pages} {1578--1583} (\bibinfo {year}
  {2020})}\BibitemShut {NoStop}%
\bibitem [{\citenamefont {Pavošević}, \citenamefont {Tao},\ and\
  \citenamefont {Hammes-Schiffer}(2021)}]{pavosevic2021dfccsd}%
  \BibitemOpen
  \bibfield  {author} {\bibinfo {author} {\bibfnamefont {F.}~\bibnamefont
  {Pavošević}}, \bibinfo {author} {\bibfnamefont {Z.}~\bibnamefont {Tao}}, \
  and\ \bibinfo {author} {\bibfnamefont {S.}~\bibnamefont {Hammes-Schiffer}},\
  }\bibfield  {title} {\enquote {\bibinfo {title} {Multicomponent coupled
  cluster singles and doubles with density fitting: Protonated water tetramers
  with quantized protons},}\ }\href@noop {} {\bibfield  {journal} {\bibinfo
  {journal} {J. Phys. Chem. Lett.}\ }\textbf {\bibinfo {volume} {12}},\
  \bibinfo {pages} {1631--1637} (\bibinfo {year} {2021})}\BibitemShut {NoStop}%
\bibitem [{\citenamefont {Fajen}\ and\ \citenamefont
  {Brorsen}(2020)}]{fajen2020separation}%
  \BibitemOpen
  \bibfield  {author} {\bibinfo {author} {\bibfnamefont {O.~J.}\ \bibnamefont
  {Fajen}}\ and\ \bibinfo {author} {\bibfnamefont {K.~R.}\ \bibnamefont
  {Brorsen}},\ }\bibfield  {title} {\enquote {\bibinfo {title} {Separation of
  electron--electron and electron--proton correlation in multicomponent
  orbital-optimized perturbation theory},}\ }\href@noop {} {\bibfield
  {journal} {\bibinfo  {journal} {J. Chem. Phys.}\ }\textbf {\bibinfo {volume}
  {152}},\ \bibinfo {pages} {194107} (\bibinfo {year} {2020})}\BibitemShut
  {NoStop}%
\bibitem [{\citenamefont {Fajen}\ and\ \citenamefont
  {Brorsen}(2021{\natexlab{a}})}]{fajen2021multicomponent}%
  \BibitemOpen
  \bibfield  {author} {\bibinfo {author} {\bibfnamefont {O.~J.}\ \bibnamefont
  {Fajen}}\ and\ \bibinfo {author} {\bibfnamefont {K.~R.}\ \bibnamefont
  {Brorsen}},\ }\bibfield  {title} {\enquote {\bibinfo {title} {Multicomponent
  casscf revisited: Large active spaces are needed for qualitatively accurate
  protonic densities},}\ }\href@noop {} {\bibfield  {journal} {\bibinfo
  {journal} {J. Chem. Theory Comput.}\ }\textbf {\bibinfo {volume} {17}},\
  \bibinfo {pages} {965--974} (\bibinfo {year}
  {2021}{\natexlab{a}})}\BibitemShut {NoStop}%
\bibitem [{\citenamefont {Hermann}, \citenamefont {DiStasio},\ and\
  \citenamefont {Tkatchenko}(2017)}]{hermann2017}%
  \BibitemOpen
  \bibfield  {author} {\bibinfo {author} {\bibfnamefont {J.}~\bibnamefont
  {Hermann}}, \bibinfo {author} {\bibfnamefont {R.~A.}\ \bibnamefont
  {DiStasio}}, \ and\ \bibinfo {author} {\bibfnamefont {A.}~\bibnamefont
  {Tkatchenko}},\ }\bibfield  {title} {\enquote {\bibinfo {title}
  {First-principles models for van der waals interactions in molecules and
  materials: Concepts, theory, and applications},}\ }\href@noop {} {\bibfield
  {journal} {\bibinfo  {journal} {Chem. Rev.}\ }\textbf {\bibinfo {volume}
  {117}},\ \bibinfo {pages} {4714--4758} (\bibinfo {year} {2017})}\BibitemShut
  {NoStop}%
\bibitem [{\citenamefont {Cohen}, \citenamefont {Mori-S{\'a}nchez},\ and\
  \citenamefont {Yang}(2008)}]{cohen2008insights}%
  \BibitemOpen
  \bibfield  {author} {\bibinfo {author} {\bibfnamefont {A.~J.}\ \bibnamefont
  {Cohen}}, \bibinfo {author} {\bibfnamefont {P.}~\bibnamefont
  {Mori-S{\'a}nchez}}, \ and\ \bibinfo {author} {\bibfnamefont
  {W.}~\bibnamefont {Yang}},\ }\bibfield  {title} {\enquote {\bibinfo {title}
  {Insights into current limitations of density functional theory},}\
  }\href@noop {} {\bibfield  {journal} {\bibinfo  {journal} {Science}\ }\textbf
  {\bibinfo {volume} {321}},\ \bibinfo {pages} {792--794} (\bibinfo {year}
  {2008})}\BibitemShut {NoStop}%
\bibitem [{\citenamefont {Ellis}, \citenamefont {Aggarwal},\ and\ \citenamefont
  {Chakraborty}(2016)}]{ellis2016development}%
  \BibitemOpen
  \bibfield  {author} {\bibinfo {author} {\bibfnamefont {B.~H.}\ \bibnamefont
  {Ellis}}, \bibinfo {author} {\bibfnamefont {S.}~\bibnamefont {Aggarwal}}, \
  and\ \bibinfo {author} {\bibfnamefont {A.}~\bibnamefont {Chakraborty}},\
  }\bibfield  {title} {\enquote {\bibinfo {title} {Development of the
  multicomponent coupled-cluster theory for investigation of multiexcitonic
  interactions},}\ }\href@noop {} {\bibfield  {journal} {\bibinfo  {journal}
  {J. Chem. Theory Comput.}\ }\textbf {\bibinfo {volume} {12}},\ \bibinfo
  {pages} {188--200} (\bibinfo {year} {2016})}\BibitemShut {NoStop}%
\bibitem [{\citenamefont {Pavo{\v{s}}evi{\'c}}\ and\ \citenamefont
  {Hammes-Schiffer}(2021)}]{pavosevic2021multicomponent}%
  \BibitemOpen
  \bibfield  {author} {\bibinfo {author} {\bibfnamefont {F.}~\bibnamefont
  {Pavo{\v{s}}evi{\'c}}}\ and\ \bibinfo {author} {\bibfnamefont
  {S.}~\bibnamefont {Hammes-Schiffer}},\ }\bibfield  {title} {\enquote
  {\bibinfo {title} {Multicomponent unitary coupled cluster and
  equation-of-motion for quantum computation.}}\ }\href@noop {} {\bibfield
  {journal} {\bibinfo  {journal} {J. Chem. Theory Comput.}\ }\textbf {\bibinfo
  {volume} {17}},\ \bibinfo {pages} {3252–3258} (\bibinfo {year}
  {2021})}\BibitemShut {NoStop}%
\bibitem [{\citenamefont {Crawford}\ and\ \citenamefont
  {Schaefer}(2000)}]{crawford2000introduction}%
  \BibitemOpen
  \bibfield  {author} {\bibinfo {author} {\bibfnamefont {T.~D.}\ \bibnamefont
  {Crawford}}\ and\ \bibinfo {author} {\bibfnamefont {H.~F.}\ \bibnamefont
  {Schaefer}},\ }\bibfield  {title} {\enquote {\bibinfo {title} {An
  introduction to coupled cluster theory for computational chemists},}\
  }\href@noop {} {\bibfield  {journal} {\bibinfo  {journal} {Reviews in
  computational chemistry}\ }\textbf {\bibinfo {volume} {14}},\ \bibinfo
  {pages} {33--136} (\bibinfo {year} {2000})}\BibitemShut {NoStop}%
\bibitem [{\citenamefont {Bartlett}\ and\ \citenamefont
  {Musia{\l}}(2007)}]{bartlett2007coupled}%
  \BibitemOpen
  \bibfield  {author} {\bibinfo {author} {\bibfnamefont {R.~J.}\ \bibnamefont
  {Bartlett}}\ and\ \bibinfo {author} {\bibfnamefont {M.}~\bibnamefont
  {Musia{\l}}},\ }\bibfield  {title} {\enquote {\bibinfo {title}
  {Coupled-cluster theory in quantum chemistry},}\ }\href@noop {} {\bibfield
  {journal} {\bibinfo  {journal} {Rev. Mod. Phys.}\ }\textbf {\bibinfo {volume}
  {79}},\ \bibinfo {pages} {291} (\bibinfo {year} {2007})}\BibitemShut
  {NoStop}%
\bibitem [{\citenamefont {Shavitt}\ and\ \citenamefont
  {Bartlett}(2009)}]{shavitt2009many}%
  \BibitemOpen
  \bibfield  {author} {\bibinfo {author} {\bibfnamefont {I.}~\bibnamefont
  {Shavitt}}\ and\ \bibinfo {author} {\bibfnamefont {R.~J.}\ \bibnamefont
  {Bartlett}},\ }\href@noop {} {\emph {\bibinfo {title} {Many-Body Methods in
  Chemistry and Physics: MBPT and Coupled-Cluster Theory}}}\ (\bibinfo
  {publisher} {Cambridge University Press},\ \bibinfo {year}
  {2009})\BibitemShut {NoStop}%
\bibitem [{\citenamefont {Fetherolf}\ \emph {et~al.}(2022)\citenamefont
  {Fetherolf}, \citenamefont {Pavo\v{s}evi\'{c}}, \citenamefont {Tao},\ and\
  \citenamefont {Hammes-Schiffer}}]{fetherolf_oomp2}%
  \BibitemOpen
  \bibfield  {author} {\bibinfo {author} {\bibfnamefont {J.~H.}\ \bibnamefont
  {Fetherolf}}, \bibinfo {author} {\bibfnamefont {F.}~\bibnamefont
  {Pavo\v{s}evi\'{c}}}, \bibinfo {author} {\bibfnamefont {Z.}~\bibnamefont
  {Tao}}, \ and\ \bibinfo {author} {\bibfnamefont {S.}~\bibnamefont
  {Hammes-Schiffer}},\ }\bibfield  {title} {\enquote {\bibinfo {title}
  {Multicomponent orbital-optimized perturbation theory with density fitting:
  Anharmonic zero-point energies in protonated water clusters},}\ }\href@noop
  {} {\bibfield  {journal} {\bibinfo  {journal} {J. Phys. Chem. Lett.}\
  }\textbf {\bibinfo {volume} {13}},\ \bibinfo {pages} {5563–5570} (\bibinfo
  {year} {2022})}\BibitemShut {NoStop}%
\bibitem [{\citenamefont {Fajen}\ and\ \citenamefont
  {Brorsen}(2021{\natexlab{b}})}]{fajen2021mp4}%
  \BibitemOpen
  \bibfield  {author} {\bibinfo {author} {\bibfnamefont {O.~J.}\ \bibnamefont
  {Fajen}}\ and\ \bibinfo {author} {\bibfnamefont {K.~R.}\ \bibnamefont
  {Brorsen}},\ }\bibfield  {title} {\enquote {\bibinfo {title} {Multicomponent
  mp4 and the inclusion of triple excitations in multicomponent many-body
  methods},}\ }\href@noop {} {\bibfield  {journal} {\bibinfo  {journal} {J.
  Chem. Phys.}\ }\textbf {\bibinfo {volume} {155}},\ \bibinfo {pages} {234108}
  (\bibinfo {year} {2021}{\natexlab{b}})}\BibitemShut {NoStop}%
\bibitem [{\citenamefont {Christiansen}, \citenamefont {Koch},\ and\
  \citenamefont {J{\o}rgensen}()}]{christiansen1995second}%
  \BibitemOpen
  \bibfield  {author} {\bibinfo {author} {\bibfnamefont {O.}~\bibnamefont
  {Christiansen}}, \bibinfo {author} {\bibfnamefont {H.}~\bibnamefont {Koch}},
  \ and\ \bibinfo {author} {\bibfnamefont {P.}~\bibnamefont {J{\o}rgensen}},\
  }\bibfield  {title} {\enquote {\bibinfo {title} {The second-order approximate
  coupled cluster singles and doubles model cc2},}\ }\href@noop {} {\bibinfo
  {journal} {Chem. Phys. Lett.}\ }\BibitemShut {NoStop}%
\bibitem [{\citenamefont {Hellweg}, \citenamefont {Gr{\"u}n},\ and\
  \citenamefont {H{\"a}ttig}(2008)}]{hellweg2008benchmarking}%
  \BibitemOpen
\bibfield  {journal} {  }\bibfield  {author} {\bibinfo {author} {\bibfnamefont
  {A.}~\bibnamefont {Hellweg}}, \bibinfo {author} {\bibfnamefont {S.~A.}\
  \bibnamefont {Gr{\"u}n}}, \ and\ \bibinfo {author} {\bibfnamefont
  {C.}~\bibnamefont {H{\"a}ttig}},\ }\bibfield  {title} {\enquote {\bibinfo
  {title} {Benchmarking the performance of spin-component scaled cc2 in ground
  and electronically excited states},}\ }\href@noop {} {\bibfield  {journal}
  {\bibinfo  {journal} {Phys. Chem. Chem. Phys.}\ }\textbf {\bibinfo {volume}
  {10}},\ \bibinfo {pages} {4119--4127} (\bibinfo {year} {2008})}\BibitemShut
  {NoStop}%
\bibitem [{\citenamefont {Tajti}\ and\ \citenamefont
  {Szalay}(2019)}]{tajti2019accuracy}%
  \BibitemOpen
  \bibfield  {author} {\bibinfo {author} {\bibfnamefont {A.}~\bibnamefont
  {Tajti}}\ and\ \bibinfo {author} {\bibfnamefont {P.~G.}\ \bibnamefont
  {Szalay}},\ }\bibfield  {title} {\enquote {\bibinfo {title} {Accuracy of
  spin-component-scaled cc2 excitation energies and potential energy
  surfaces},}\ }\href@noop {} {\bibfield  {journal} {\bibinfo  {journal} {J.
  Chem. Theory Comput.}\ }\textbf {\bibinfo {volume} {15}},\ \bibinfo {pages}
  {5523--5531} (\bibinfo {year} {2019})}\BibitemShut {NoStop}%
\bibitem [{\citenamefont {Pavo{\v{s}}evi{\'c}}\ and\ \citenamefont
  {Hammes-Schiffer}(2019{\natexlab{b}})}]{pavovsevic2019eomccsd}%
  \BibitemOpen
  \bibfield  {author} {\bibinfo {author} {\bibfnamefont {F.}~\bibnamefont
  {Pavo{\v{s}}evi{\'c}}}\ and\ \bibinfo {author} {\bibfnamefont
  {S.}~\bibnamefont {Hammes-Schiffer}},\ }\bibfield  {title} {\enquote
  {\bibinfo {title} {Multicomponent equation-of-motion coupled cluster singles
  and doubles: Theory and calculation of excitation energies for positronium
  hydride},}\ }\href@noop {} {\bibfield  {journal} {\bibinfo  {journal} {J.
  Chem. Phys.}\ }\textbf {\bibinfo {volume} {150}},\ \bibinfo {pages} {161102}
  (\bibinfo {year} {2019}{\natexlab{b}})}\BibitemShut {NoStop}%
\bibitem [{\citenamefont {Pavošević}\ \emph {et~al.}(2020)\citenamefont
  {Pavošević}, \citenamefont {Tao}, \citenamefont {Culpitt}, \citenamefont
  {Zhao}, \citenamefont {Li},\ and\ \citenamefont
  {Hammes-Schiffer}}]{pavosevic2020frequency}%
  \BibitemOpen
  \bibfield  {author} {\bibinfo {author} {\bibfnamefont {F.}~\bibnamefont
  {Pavošević}}, \bibinfo {author} {\bibfnamefont {Z.}~\bibnamefont {Tao}},
  \bibinfo {author} {\bibfnamefont {T.}~\bibnamefont {Culpitt}}, \bibinfo
  {author} {\bibfnamefont {L.}~\bibnamefont {Zhao}}, \bibinfo {author}
  {\bibfnamefont {X.}~\bibnamefont {Li}}, \ and\ \bibinfo {author}
  {\bibfnamefont {S.}~\bibnamefont {Hammes-Schiffer}},\ }\bibfield  {title}
  {\enquote {\bibinfo {title} {Frequency and time domain nuclear--electronic
  orbital equation-of-motion coupled cluster methods: Combination bands and
  electronic--protonic double excitations},}\ }\href@noop {} {\bibfield
  {journal} {\bibinfo  {journal} {J. Phys. Chem. Lett.}\ }\textbf {\bibinfo
  {volume} {11}},\ \bibinfo {pages} {6435--6442} (\bibinfo {year}
  {2020})}\BibitemShut {NoStop}%
\bibitem [{\citenamefont {Pavo{\v{s}}evi{\'c}}\ and\ \citenamefont
  {Hammes-Schiffer}(2020)}]{pavovsevic2020automatic}%
  \BibitemOpen
  \bibfield  {author} {\bibinfo {author} {\bibfnamefont {F.}~\bibnamefont
  {Pavo{\v{s}}evi{\'c}}}\ and\ \bibinfo {author} {\bibfnamefont
  {S.}~\bibnamefont {Hammes-Schiffer}},\ }\bibfield  {title} {\enquote
  {\bibinfo {title} {Automatic differentiation for coupled cluster methods},}\
  }\href@noop {} {\bibfield  {journal} {\bibinfo  {journal} {arXiv preprint
  arXiv:2011.11690}\ } (\bibinfo {year} {2020})}\BibitemShut {NoStop}%
\bibitem [{\citenamefont {Kutzelnigg}\ and\ \citenamefont
  {Mukherjee}(1997)}]{kutzelnigg1997normal}%
  \BibitemOpen
  \bibfield  {author} {\bibinfo {author} {\bibfnamefont {W.}~\bibnamefont
  {Kutzelnigg}}\ and\ \bibinfo {author} {\bibfnamefont {D.}~\bibnamefont
  {Mukherjee}},\ }\bibfield  {title} {\enquote {\bibinfo {title} {Normal order
  and extended wick theorem for a multiconfiguration reference wave
  function},}\ }\href@noop {} {\bibfield  {journal} {\bibinfo  {journal} {J.
  Chem. Phys.}\ }\textbf {\bibinfo {volume} {107}},\ \bibinfo {pages}
  {432--449} (\bibinfo {year} {1997})}\BibitemShut {NoStop}%
\bibitem [{\citenamefont {Neese}\ \emph {et~al.}(2009)\citenamefont {Neese},
  \citenamefont {Schwabe}, \citenamefont {Kossmann}, \citenamefont {Schirmer},\
  and\ \citenamefont {Grimme}}]{neese2009assessment}%
  \BibitemOpen
  \bibfield  {author} {\bibinfo {author} {\bibfnamefont {F.}~\bibnamefont
  {Neese}}, \bibinfo {author} {\bibfnamefont {T.}~\bibnamefont {Schwabe}},
  \bibinfo {author} {\bibfnamefont {S.}~\bibnamefont {Kossmann}}, \bibinfo
  {author} {\bibfnamefont {B.}~\bibnamefont {Schirmer}}, \ and\ \bibinfo
  {author} {\bibfnamefont {S.}~\bibnamefont {Grimme}},\ }\bibfield  {title}
  {\enquote {\bibinfo {title} {Assessment of orbital-optimized, spin-component
  scaled second-order many-body perturbation theory for thermochemistry and
  kinetics},}\ }\href@noop {} {\bibfield  {journal} {\bibinfo  {journal} {J.
  Chem. Theory Comput.}\ }\textbf {\bibinfo {volume} {5}},\ \bibinfo {pages}
  {3060--3073} (\bibinfo {year} {2009})}\BibitemShut {NoStop}%
\bibitem [{\citenamefont {Jung}\ \emph {et~al.}(2004)\citenamefont {Jung},
  \citenamefont {Lochan}, \citenamefont {Dutoi},\ and\ \citenamefont
  {Head-Gordon}}]{jung2004scaled}%
  \BibitemOpen
  \bibfield  {author} {\bibinfo {author} {\bibfnamefont {Y.}~\bibnamefont
  {Jung}}, \bibinfo {author} {\bibfnamefont {R.~C.}\ \bibnamefont {Lochan}},
  \bibinfo {author} {\bibfnamefont {A.~D.}\ \bibnamefont {Dutoi}}, \ and\
  \bibinfo {author} {\bibfnamefont {M.}~\bibnamefont {Head-Gordon}},\
  }\bibfield  {title} {\enquote {\bibinfo {title} {Scaled opposite-spin second
  order m{\o}ller--plesset correlation energy: An economical electronic
  structure method},}\ }\href@noop {} {\bibfield  {journal} {\bibinfo
  {journal} {J. Chem. Phys.}\ }\textbf {\bibinfo {volume} {121}},\ \bibinfo
  {pages} {9793--9802} (\bibinfo {year} {2004})}\BibitemShut {NoStop}%
\bibitem [{\citenamefont {Distasio~Jr}\ and\ \citenamefont
  {Head-Gordon}(2007)}]{distasio2007optimized}%
  \BibitemOpen
  \bibfield  {author} {\bibinfo {author} {\bibfnamefont {R.~A.}\ \bibnamefont
  {Distasio~Jr}}\ and\ \bibinfo {author} {\bibfnamefont {M.}~\bibnamefont
  {Head-Gordon}},\ }\bibfield  {title} {\enquote {\bibinfo {title} {Optimized
  spin-component scaled second-order m{\o}ller-plesset perturbation theory for
  intermolecular interaction energies},}\ }\href@noop {} {\bibfield  {journal}
  {\bibinfo  {journal} {Mol. Phys.}\ }\textbf {\bibinfo {volume} {105}},\
  \bibinfo {pages} {1073--1083} (\bibinfo {year} {2007})}\BibitemShut {NoStop}%
\bibitem [{\citenamefont {Whitten}(1973)}]{whitten1973coulombic}%
  \BibitemOpen
  \bibfield  {author} {\bibinfo {author} {\bibfnamefont {J.~L.}\ \bibnamefont
  {Whitten}},\ }\bibfield  {title} {\enquote {\bibinfo {title} {Coulombic
  potential energy integrals and approximations},}\ }\href@noop {} {\bibfield
  {journal} {\bibinfo  {journal} {J. Chem. Phys.}\ }\textbf {\bibinfo {volume}
  {58}},\ \bibinfo {pages} {4496--4501} (\bibinfo {year} {1973})}\BibitemShut
  {NoStop}%
\bibitem [{\citenamefont {Dunlap}, \citenamefont {Connolly},\ and\
  \citenamefont {Sabin}(1979)}]{dunlap1979some}%
  \BibitemOpen
  \bibfield  {author} {\bibinfo {author} {\bibfnamefont {B.~I.}\ \bibnamefont
  {Dunlap}}, \bibinfo {author} {\bibfnamefont {J.}~\bibnamefont {Connolly}}, \
  and\ \bibinfo {author} {\bibfnamefont {J.}~\bibnamefont {Sabin}},\ }\bibfield
   {title} {\enquote {\bibinfo {title} {On some approximations in applications
  of x $\alpha$ theory},}\ }\href@noop {} {\bibfield  {journal} {\bibinfo
  {journal} {J. Chem. Phys.}\ }\textbf {\bibinfo {volume} {71}},\ \bibinfo
  {pages} {3396--3402} (\bibinfo {year} {1979})}\BibitemShut {NoStop}%
\bibitem [{\citenamefont {Mej{\'\i}a-Rodr{\'\i}guez}\ and\ \citenamefont {de~la
  Lande}(2019)}]{mejia2019multicomponent}%
  \BibitemOpen
  \bibfield  {author} {\bibinfo {author} {\bibfnamefont {D.}~\bibnamefont
  {Mej{\'\i}a-Rodr{\'\i}guez}}\ and\ \bibinfo {author} {\bibfnamefont
  {A.}~\bibnamefont {de~la Lande}},\ }\bibfield  {title} {\enquote {\bibinfo
  {title} {Multicomponent density functional theory with density fitting},}\
  }\href@noop {} {\bibfield  {journal} {\bibinfo  {journal} {J. Chem. Phys.}\
  }\textbf {\bibinfo {volume} {150}},\ \bibinfo {pages} {174115} (\bibinfo
  {year} {2019})}\BibitemShut {NoStop}%
\bibitem [{\citenamefont {Smith}\ \emph {et~al.}(2018)\citenamefont {Smith},
  \citenamefont {Burns}, \citenamefont {Sirianni}, \citenamefont {Nascimento},
  \citenamefont {Kumar}, \citenamefont {James}, \citenamefont {Schriber},
  \citenamefont {Zhang}, \citenamefont {Zhang}, \citenamefont {Abbott} \emph
  {et~al.}}]{smith2018psi4numpy}%
  \BibitemOpen
  \bibfield  {author} {\bibinfo {author} {\bibfnamefont {D.~G.}\ \bibnamefont
  {Smith}}, \bibinfo {author} {\bibfnamefont {L.~A.}\ \bibnamefont {Burns}},
  \bibinfo {author} {\bibfnamefont {D.~A.}\ \bibnamefont {Sirianni}}, \bibinfo
  {author} {\bibfnamefont {D.~R.}\ \bibnamefont {Nascimento}}, \bibinfo
  {author} {\bibfnamefont {A.}~\bibnamefont {Kumar}}, \bibinfo {author}
  {\bibfnamefont {A.~M.}\ \bibnamefont {James}}, \bibinfo {author}
  {\bibfnamefont {J.~B.}\ \bibnamefont {Schriber}}, \bibinfo {author}
  {\bibfnamefont {T.}~\bibnamefont {Zhang}}, \bibinfo {author} {\bibfnamefont
  {B.}~\bibnamefont {Zhang}}, \bibinfo {author} {\bibfnamefont {A.~S.}\
  \bibnamefont {Abbott}},  \emph {et~al.},\ }\bibfield  {title} {\enquote
  {\bibinfo {title} {Psi4numpy: An interactive quantum chemistry programming
  environment for reference implementations and rapid development},}\
  }\href@noop {} {\bibfield  {journal} {\bibinfo  {journal} {J. Chem. Theory
  Comput.}\ }\textbf {\bibinfo {volume} {14}},\ \bibinfo {pages} {3504--3511}
  (\bibinfo {year} {2018})}\BibitemShut {NoStop}%
\bibitem [{\citenamefont {Valeev}(2014)}]{githubsequant}%
  \BibitemOpen
  \bibfield  {author} {\bibinfo {author} {\bibfnamefont {E.~F.}\ \bibnamefont
  {Valeev}},\ }\href@noop {} {}\bibinfo {howpublished}
  {https://github.com/ValeevGroup/SeQuant} (\bibinfo {year} {2014})\BibitemShut
  {NoStop}%
\bibitem [{\citenamefont {Abadi}\ \emph {et~al.}(2015)\citenamefont {Abadi},
  \citenamefont {Agarwal}, \citenamefont {Barham}, \citenamefont {Brevdo},
  \citenamefont {Chen}, \citenamefont {Citro}, \citenamefont {Corrado},
  \citenamefont {Davis}, \citenamefont {Dean}, \citenamefont {Devin},
  \citenamefont {Ghemawat}, \citenamefont {Goodfellow}, \citenamefont {Harp},
  \citenamefont {Irving}, \citenamefont {Isard}, \citenamefont {Jia},
  \citenamefont {Jozefowicz}, \citenamefont {Kaiser}, \citenamefont {Kudlur},
  \citenamefont {Levenberg}, \citenamefont {Man\'{e}}, \citenamefont {Monga},
  \citenamefont {Moore}, \citenamefont {Murray}, \citenamefont {Olah},
  \citenamefont {Schuster}, \citenamefont {Shlens}, \citenamefont {Steiner},
  \citenamefont {Sutskever}, \citenamefont {Talwar}, \citenamefont {Tucker},
  \citenamefont {Vanhoucke}, \citenamefont {Vasudevan}, \citenamefont
  {Vi\'{e}gas}, \citenamefont {Vinyals}, \citenamefont {Warden}, \citenamefont
  {Wattenberg}, \citenamefont {Wicke}, \citenamefont {Yu},\ and\ \citenamefont
  {Zheng}}]{tensorflow2015}%
  \BibitemOpen
  \bibfield  {author} {\bibinfo {author} {\bibfnamefont {M.}~\bibnamefont
  {Abadi}}, \bibinfo {author} {\bibfnamefont {A.}~\bibnamefont {Agarwal}},
  \bibinfo {author} {\bibfnamefont {P.}~\bibnamefont {Barham}}, \bibinfo
  {author} {\bibfnamefont {E.}~\bibnamefont {Brevdo}}, \bibinfo {author}
  {\bibfnamefont {Z.}~\bibnamefont {Chen}}, \bibinfo {author} {\bibfnamefont
  {C.}~\bibnamefont {Citro}}, \bibinfo {author} {\bibfnamefont {G.~S.}\
  \bibnamefont {Corrado}}, \bibinfo {author} {\bibfnamefont {A.}~\bibnamefont
  {Davis}}, \bibinfo {author} {\bibfnamefont {J.}~\bibnamefont {Dean}},
  \bibinfo {author} {\bibfnamefont {M.}~\bibnamefont {Devin}}, \bibinfo
  {author} {\bibfnamefont {S.}~\bibnamefont {Ghemawat}}, \bibinfo {author}
  {\bibfnamefont {I.}~\bibnamefont {Goodfellow}}, \bibinfo {author}
  {\bibfnamefont {A.}~\bibnamefont {Harp}}, \bibinfo {author} {\bibfnamefont
  {G.}~\bibnamefont {Irving}}, \bibinfo {author} {\bibfnamefont
  {M.}~\bibnamefont {Isard}}, \bibinfo {author} {\bibfnamefont
  {Y.}~\bibnamefont {Jia}}, \bibinfo {author} {\bibfnamefont {R.}~\bibnamefont
  {Jozefowicz}}, \bibinfo {author} {\bibfnamefont {L.}~\bibnamefont {Kaiser}},
  \bibinfo {author} {\bibfnamefont {M.}~\bibnamefont {Kudlur}}, \bibinfo
  {author} {\bibfnamefont {J.}~\bibnamefont {Levenberg}}, \bibinfo {author}
  {\bibfnamefont {D.}~\bibnamefont {Man\'{e}}}, \bibinfo {author}
  {\bibfnamefont {R.}~\bibnamefont {Monga}}, \bibinfo {author} {\bibfnamefont
  {S.}~\bibnamefont {Moore}}, \bibinfo {author} {\bibfnamefont
  {D.}~\bibnamefont {Murray}}, \bibinfo {author} {\bibfnamefont
  {C.}~\bibnamefont {Olah}}, \bibinfo {author} {\bibfnamefont {M.}~\bibnamefont
  {Schuster}}, \bibinfo {author} {\bibfnamefont {J.}~\bibnamefont {Shlens}},
  \bibinfo {author} {\bibfnamefont {B.}~\bibnamefont {Steiner}}, \bibinfo
  {author} {\bibfnamefont {I.}~\bibnamefont {Sutskever}}, \bibinfo {author}
  {\bibfnamefont {K.}~\bibnamefont {Talwar}}, \bibinfo {author} {\bibfnamefont
  {P.}~\bibnamefont {Tucker}}, \bibinfo {author} {\bibfnamefont
  {V.}~\bibnamefont {Vanhoucke}}, \bibinfo {author} {\bibfnamefont
  {V.}~\bibnamefont {Vasudevan}}, \bibinfo {author} {\bibfnamefont
  {F.}~\bibnamefont {Vi\'{e}gas}}, \bibinfo {author} {\bibfnamefont
  {O.}~\bibnamefont {Vinyals}}, \bibinfo {author} {\bibfnamefont
  {P.}~\bibnamefont {Warden}}, \bibinfo {author} {\bibfnamefont
  {M.}~\bibnamefont {Wattenberg}}, \bibinfo {author} {\bibfnamefont
  {M.}~\bibnamefont {Wicke}}, \bibinfo {author} {\bibfnamefont
  {Y.}~\bibnamefont {Yu}}, \ and\ \bibinfo {author} {\bibfnamefont
  {X.}~\bibnamefont {Zheng}},\ }\href {https://www.tensorflow.org/} {\enquote
  {\bibinfo {title} {{TensorFlow}: Large-scale machine learning on
  heterogeneous systems},}\ } (\bibinfo {year} {2015}),\ \bibinfo {note}
  {software available from tensorflow.org}\BibitemShut {NoStop}%
\bibitem [{\citenamefont {Dunning~Jr}(1989)}]{dunning1989gaussian}%
  \BibitemOpen
  \bibfield  {author} {\bibinfo {author} {\bibfnamefont {T.~H.}\ \bibnamefont
  {Dunning~Jr}},\ }\bibfield  {title} {\enquote {\bibinfo {title} {Gaussian
  basis sets for use in correlated molecular calculations. i. the atoms boron
  through neon and hydrogen},}\ }\href@noop {} {\bibfield  {journal} {\bibinfo
  {journal} {J. Chem. Phys.}\ }\textbf {\bibinfo {volume} {90}},\ \bibinfo
  {pages} {1007--1023} (\bibinfo {year} {1989})}\BibitemShut {NoStop}%
\bibitem [{\citenamefont {Kendall}, \citenamefont {Dunning~Jr},\ and\
  \citenamefont {Harrison}(1992)}]{kendall1992electron}%
  \BibitemOpen
  \bibfield  {author} {\bibinfo {author} {\bibfnamefont {R.~A.}\ \bibnamefont
  {Kendall}}, \bibinfo {author} {\bibfnamefont {T.~H.}\ \bibnamefont
  {Dunning~Jr}}, \ and\ \bibinfo {author} {\bibfnamefont {R.~J.}\ \bibnamefont
  {Harrison}},\ }\bibfield  {title} {\enquote {\bibinfo {title} {Electron
  affinities of the first-row atoms revisited. systematic basis sets and wave
  functions},}\ }\href@noop {} {\bibfield  {journal} {\bibinfo  {journal} {J.
  Chem. Phys.}\ }\textbf {\bibinfo {volume} {96}},\ \bibinfo {pages}
  {6796--6806} (\bibinfo {year} {1992})}\BibitemShut {NoStop}%
\bibitem [{\citenamefont {Weigend}\ \emph {et~al.}(1998)\citenamefont
  {Weigend}, \citenamefont {H{\"a}ser}, \citenamefont {Patzelt},\ and\
  \citenamefont {Ahlrichs}}]{weigend1998ri}%
  \BibitemOpen
  \bibfield  {author} {\bibinfo {author} {\bibfnamefont {F.}~\bibnamefont
  {Weigend}}, \bibinfo {author} {\bibfnamefont {M.}~\bibnamefont {H{\"a}ser}},
  \bibinfo {author} {\bibfnamefont {H.}~\bibnamefont {Patzelt}}, \ and\
  \bibinfo {author} {\bibfnamefont {R.}~\bibnamefont {Ahlrichs}},\ }\bibfield
  {title} {\enquote {\bibinfo {title} {Ri-mp2: optimized auxiliary basis sets
  and demonstration of efficiency},}\ }\href@noop {} {\bibfield  {journal}
  {\bibinfo  {journal} {Chem. Phys. Lett.}\ }\textbf {\bibinfo {volume}
  {294}},\ \bibinfo {pages} {143--152} (\bibinfo {year} {1998})}\BibitemShut
  {NoStop}%
\bibitem [{\citenamefont {Weigend}, \citenamefont {K{\"o}hn},\ and\
  \citenamefont {H{\"a}ttig}(2002)}]{weigend2002efficient}%
  \BibitemOpen
  \bibfield  {author} {\bibinfo {author} {\bibfnamefont {F.}~\bibnamefont
  {Weigend}}, \bibinfo {author} {\bibfnamefont {A.}~\bibnamefont {K{\"o}hn}}, \
  and\ \bibinfo {author} {\bibfnamefont {C.}~\bibnamefont {H{\"a}ttig}},\
  }\bibfield  {title} {\enquote {\bibinfo {title} {Efficient use of the
  correlation consistent basis sets in resolution of the identity mp2
  calculations},}\ }\href@noop {} {\bibfield  {journal} {\bibinfo  {journal}
  {J. Chem. Phys.}\ }\textbf {\bibinfo {volume} {116}},\ \bibinfo {pages}
  {3175--3183} (\bibinfo {year} {2002})}\BibitemShut {NoStop}%
\bibitem [{\citenamefont {Yu}, \citenamefont {Pavo{\v{s}}evi{\'c}},\ and\
  \citenamefont {Hammes-Schiffer}(2020)}]{yu2020development}%
  \BibitemOpen
  \bibfield  {author} {\bibinfo {author} {\bibfnamefont {Q.}~\bibnamefont
  {Yu}}, \bibinfo {author} {\bibfnamefont {F.}~\bibnamefont
  {Pavo{\v{s}}evi{\'c}}}, \ and\ \bibinfo {author} {\bibfnamefont
  {S.}~\bibnamefont {Hammes-Schiffer}},\ }\bibfield  {title} {\enquote
  {\bibinfo {title} {Development of nuclear basis sets for multicomponent
  quantum chemistry methods},}\ }\href@noop {} {\bibfield  {journal} {\bibinfo
  {journal} {J. Chem. Phys.}\ }\textbf {\bibinfo {volume} {152}},\ \bibinfo
  {pages} {244123} (\bibinfo {year} {2020})}\BibitemShut {NoStop}%
\bibitem [{\citenamefont {Culpitt}\ \emph {et~al.}(2019)\citenamefont
  {Culpitt}, \citenamefont {Yang}, \citenamefont {Pavo{\v{s}}evi{\'c}},
  \citenamefont {Tao},\ and\ \citenamefont
  {Hammes-Schiffer}}]{culpitt2019enhancing}%
  \BibitemOpen
  \bibfield  {author} {\bibinfo {author} {\bibfnamefont {T.}~\bibnamefont
  {Culpitt}}, \bibinfo {author} {\bibfnamefont {Y.}~\bibnamefont {Yang}},
  \bibinfo {author} {\bibfnamefont {F.}~\bibnamefont {Pavo{\v{s}}evi{\'c}}},
  \bibinfo {author} {\bibfnamefont {Z.}~\bibnamefont {Tao}}, \ and\ \bibinfo
  {author} {\bibfnamefont {S.}~\bibnamefont {Hammes-Schiffer}},\ }\bibfield
  {title} {\enquote {\bibinfo {title} {Enhancing the applicability of
  multicomponent time-dependent density functional theory},}\ }\href@noop {}
  {\bibfield  {journal} {\bibinfo  {journal} {J. Chem. Phys.}\ }\textbf
  {\bibinfo {volume} {150}},\ \bibinfo {pages} {201101} (\bibinfo {year}
  {2019})}\BibitemShut {NoStop}%
\bibitem [{\citenamefont {Webb}\ and\ \citenamefont
  {Hammes-Schiffer}(2000)}]{webb2000fourier}%
  \BibitemOpen
  \bibfield  {author} {\bibinfo {author} {\bibfnamefont {S.~P.}\ \bibnamefont
  {Webb}}\ and\ \bibinfo {author} {\bibfnamefont {S.}~\bibnamefont
  {Hammes-Schiffer}},\ }\bibfield  {title} {\enquote {\bibinfo {title} {Fourier
  grid hamiltonian multiconfigurational self-consistent-field: A method to
  calculate multidimensional hydrogen vibrational wavefunctions},}\ }\href@noop
  {} {\bibfield  {journal} {\bibinfo  {journal} {J. Chem. Phys.}\ }\textbf
  {\bibinfo {volume} {113}},\ \bibinfo {pages} {5214--5227} (\bibinfo {year}
  {2000})}\BibitemShut {NoStop}%
\bibitem [{\citenamefont {Cumming}\ and\ \citenamefont
  {Kebarle}(1978)}]{cumming1978summary}%
  \BibitemOpen
  \bibfield  {author} {\bibinfo {author} {\bibfnamefont {J.~B.}\ \bibnamefont
  {Cumming}}\ and\ \bibinfo {author} {\bibfnamefont {P.}~\bibnamefont
  {Kebarle}},\ }\bibfield  {title} {\enquote {\bibinfo {title} {Summary of gas
  phase acidity measurements involving acids ah. entropy changes in proton
  transfer reactions involving negative ions. bond dissociation energies d
  (a—h) and electron affinities ea (a)},}\ }\href@noop {} {\bibfield
  {journal} {\bibinfo  {journal} {Can. J. Chem.}\ }\textbf {\bibinfo {volume}
  {56}},\ \bibinfo {pages} {1--9} (\bibinfo {year} {1978})}\BibitemShut
  {NoStop}%
\bibitem [{\citenamefont {Graul}, \citenamefont {Schnute},\ and\ \citenamefont
  {Squires}(1990)}]{graul1990gas}%
  \BibitemOpen
  \bibfield  {author} {\bibinfo {author} {\bibfnamefont {S.~T.}\ \bibnamefont
  {Graul}}, \bibinfo {author} {\bibfnamefont {M.~E.}\ \bibnamefont {Schnute}},
  \ and\ \bibinfo {author} {\bibfnamefont {R.~R.}\ \bibnamefont {Squires}},\
  }\bibfield  {title} {\enquote {\bibinfo {title} {Gas-phase acidities of
  carboxylic acids and alcohols from collision-induced dissociation of dimer
  cluster ions},}\ }\href@noop {} {\bibfield  {journal} {\bibinfo  {journal}
  {Int. J. Mass Spectrom. Ion Processes}\ }\textbf {\bibinfo {volume} {96}},\
  \bibinfo {pages} {181--198} (\bibinfo {year} {1990})}\BibitemShut {NoStop}%
\bibitem [{\citenamefont {Hunter}\ and\ \citenamefont
  {Lias}(1998)}]{hunter1998evaluated}%
  \BibitemOpen
  \bibfield  {author} {\bibinfo {author} {\bibfnamefont {E.~P.}\ \bibnamefont
  {Hunter}}\ and\ \bibinfo {author} {\bibfnamefont {S.~G.}\ \bibnamefont
  {Lias}},\ }\bibfield  {title} {\enquote {\bibinfo {title} {Evaluated gas
  phase basicities and proton affinities of molecules: an update},}\
  }\href@noop {} {\bibfield  {journal} {\bibinfo  {journal} {J. Phys. Chem.
  Ref. Data}\ }\textbf {\bibinfo {volume} {27}},\ \bibinfo {pages} {413--656}
  (\bibinfo {year} {1998})}\BibitemShut {NoStop}%
\bibitem [{\citenamefont {D{\'\i}az-Tinoco}\ \emph {et~al.}(2013)\citenamefont
  {D{\'\i}az-Tinoco}, \citenamefont {Romero}, \citenamefont {Ortiz},
  \citenamefont {Reyes},\ and\ \citenamefont
  {Flores-Moreno}}]{diaz2013generalized}%
  \BibitemOpen
  \bibfield  {author} {\bibinfo {author} {\bibfnamefont {M.}~\bibnamefont
  {D{\'\i}az-Tinoco}}, \bibinfo {author} {\bibfnamefont {J.}~\bibnamefont
  {Romero}}, \bibinfo {author} {\bibfnamefont {J.}~\bibnamefont {Ortiz}},
  \bibinfo {author} {\bibfnamefont {A.}~\bibnamefont {Reyes}}, \ and\ \bibinfo
  {author} {\bibfnamefont {R.}~\bibnamefont {Flores-Moreno}},\ }\bibfield
  {title} {\enquote {\bibinfo {title} {A generalized any-particle propagator
  theory: Prediction of proton affinities and acidity properties with the
  proton propagator},}\ }\href@noop {} {\bibfield  {journal} {\bibinfo
  {journal} {J. Chem. Phys.}\ }\textbf {\bibinfo {volume} {138}},\ \bibinfo
  {pages} {194108} (\bibinfo {year} {2013})}\BibitemShut {NoStop}%
\bibitem [{\citenamefont {Kong}, \citenamefont {Bischoff},\ and\ \citenamefont
  {Valeev}(2012)}]{explicitly}%
  \BibitemOpen
  \bibfield  {author} {\bibinfo {author} {\bibfnamefont {L.}~\bibnamefont
  {Kong}}, \bibinfo {author} {\bibfnamefont {F.~A.}\ \bibnamefont {Bischoff}},
  \ and\ \bibinfo {author} {\bibfnamefont {E.~F.}\ \bibnamefont {Valeev}},\
  }\bibfield  {title} {\enquote {\bibinfo {title} {Explicitly correlated
  r12/f12 methods for electronic structure},}\ }\href@noop {} {\bibfield
  {journal} {\bibinfo  {journal} {Chem. Rev.}\ }\textbf {\bibinfo {volume}
  {112}},\ \bibinfo {pages} {75--107} (\bibinfo {year} {2012})}\BibitemShut
  {NoStop}%
\bibitem [{\citenamefont {Helgaker}\ \emph {et~al.}(1997)\citenamefont
  {Helgaker}, \citenamefont {Klopper}, \citenamefont {Koch},\ and\
  \citenamefont {Noga}}]{extrapolation}%
  \BibitemOpen
  \bibfield  {author} {\bibinfo {author} {\bibfnamefont {T.}~\bibnamefont
  {Helgaker}}, \bibinfo {author} {\bibfnamefont {W.}~\bibnamefont {Klopper}},
  \bibinfo {author} {\bibfnamefont {H.}~\bibnamefont {Koch}}, \ and\ \bibinfo
  {author} {\bibfnamefont {J.}~\bibnamefont {Noga}},\ }\bibfield  {title}
  {\enquote {\bibinfo {title} {Basis-set convergence of correlated calculations
  on water},}\ }\href@noop {} {\bibfield  {journal} {\bibinfo  {journal} {J.
  Chem. Phys.}\ }\textbf {\bibinfo {volume} {106}},\ \bibinfo {pages}
  {9639--9646} (\bibinfo {year} {1997})}\BibitemShut {NoStop}%
\end{thebibliography}%

\end{document}